\documentclass[showpacs,superscriptaddress,amsmath,amssymb,nofootinbib,twocolumn]{revtex4-2}
\usepackage{graphicx}
\usepackage{epsfig}
\usepackage{overpic}
\usepackage{dcolumn}
\usepackage{bm}
\usepackage{lineno}
\usepackage{xspace}
\usepackage{multirow}
\usepackage{epstopdf}
\usepackage{xcolor}
\usepackage{soul}
\soulregister{\cite}7
\soulregister{\ref}7
\soulregister{\st}7
\usepackage[colorlinks,allcolors=black]{hyperref}
\usepackage{verbatim}
\usepackage{enumitem}
\usepackage{todonotes}
\usepackage{subcaption}
\usepackage{dcolumn}
\usepackage{bm}
\usepackage{threeparttable}
\usepackage{multirow}
\usepackage[figuresright]{rotating}
\usepackage{makecell}
\usepackage{verbatim}
\usepackage{slashed}
\usepackage{mathtools}
\usepackage{tikz-feynman}
\usepackage{utfsym}
\usepackage{amsmath}
\include{def-com}
\newcommand{\half}{\frac12}

\newcommand{\scnt}{\affiliation{Southern Center for Nuclear-Science Theory (SCNT), Institute of Modern Physics, Chinese Academy of Sciences, Huizhou 516000, Guangdong Province, China}}
\newcommand{\OU}{\affiliation{Research Center for Nuclear Physics (RCNP), Osaka University, Ibaraki 567-0047, Japan}}

\begin{document}
\title{\boldmath Revisit the diquark of $\Lambda_c$ in the $\Lambda_c\to \Lambda K^+$ and $\Lambda_c\to \Sigma^0 K^+$ processes}

\author{Peng-Yu Niu}\email{niupy@m.scnu.edu.cn}
\affiliation{Guangdong Provincial Key Laboratory of Nuclear Science, Institute of Quantum Matter, South China Normal University, Guangzhou 510006, China}
\affiliation{Guangdong-Hong Kong Joint Laboratory of Quantum Matter, Southern Nuclear Science Computing Center, South China Normal University, Guangzhou 510006, China}

\author{Qian Wang}\email{qianwang@m.scnu.edu.cn}
\affiliation{Guangdong Provincial Key Laboratory of Nuclear Science, Institute of Quantum Matter, South China Normal University, Guangzhou 510006, China}
\affiliation{Guangdong-Hong Kong Joint Laboratory of Quantum Matter, Southern Nuclear Science Computing Center, South China Normal University, Guangzhou 510006, China}
\scnt
\OU

\author{Qiang Zhao}\email{zhaoq@ihep.ac.cn}
\affiliation{Institute of High Energy Physics, Chinese Academy of Sciences, Beijing 100049, China }
\affiliation{University of Chinese Academy of Sciences, Beijing 100049, China}
\affiliation{Center for High Energy Physics, Henan Academy of Sciences, Zhengzhou 450046, China}

\date{\today}

\begin{abstract}
The spatial distributions of $[ud]$ diquark and heavy-light diquark of the SU(3)-flavor antitriplet charmed baryons are investigated by the two singly Cabibbo-suppressed hadronic weak decays, $\Lambda_c\to \Lambda K^+$ and $\Lambda_c\to \Sigma^0 K^+$ within the nonrelativistic constituent quark model. The above two spatial distributions are reflected by the two parameters $\alpha_\rho$ and $\alpha_\lambda$, which are the
harmonic oscillator strength parameters of the charmed baryons. These two parameters obtain strong constraints from the decay widths of $\Lambda_c\to \Lambda K^+$ and $\Lambda_c\to \Sigma^0 K^+$. The larger the harmonic oscillator parameter is, the more compact the spatial distribution will become. The current $\alpha_\rho$ and $\alpha_\lambda$ indicate that neither the light diquark nor the heavy-light diquark turns out to favor a compact structure. In addition, some selection rules, including the ``$\Lambda$ selection rule'', can be useful for the search of excited baryons in the heavy-flavor baryon hadronic weak decays.

\end{abstract}

\maketitle

\section{Introduction}

Hadrons are colorless systems of (anti)quarks and gluons and have abundant structures. Those, beyond the normal meson and baryon pictures in the conventional quark model, are named as exotic hadrons. 
Tens of exotic candidates have been measured in experiment, which dramatically expand the hadron spectrum.
Their structures are one of the essential topics to understand the underlying non-perturbative properties of quantum chromodynamics (QCD). 
Due to the complicated non-perturbative mechanism, many ``effective objects", e.g. constituent quarks, are introduced to reproduce the spectrum of hadrons, as well as their decay modes. The diquark is one of the most important ``effective objects" and widely used in diverse works to understand the properties of hadrons and hadronic matter~\cite{Wilczek:2004im,Shifman:2005wa,Ali:2019roi,Yang:2020atz,Chen:2022asf,Huang:2023jec, Shifman:2024kfj,Kim:2024tbf,Liu:2019zoy,Hosaka:2016pey,Esposito:2014rxa,Klempt:2009pi,Zhu:2004xa}.

Diquark is a two-quark system in either $\bar 3$ or $6$ color irreducible representations. 
Due to the Fermi-Dirac statistics, the $\bar 3$ diquark should be spin-0 diquark, named as ``good diquarks"~\cite{Wilczek:2004im,Shifman:2005wa,Shifman:2024kfj}, and have been discussed in  
various multiquark systems ~\cite{An:2025rjv,Shi:2021jyr,Yang:2020atz,Ali:2019clg,Giannuzzi:2019esi,Maiani:2015vwa,Padmanath:2015era,Lee:2009rt,Ebert:2007rn,Karliner:2003dt,Jaffe:2003sg}.
This kind of diquark is made of two light flavor quarks with color $\bar 3$, flavor $\bar 3$, and spin $0$~\cite{Shifman:2024kfj,Kim:2024tbf}. Within such a configuration the quark-quark interaction due to the one-gluon-exchange potential is attractive and will lower the energy of the diquark system. As a consequence, the diquark may appear as a compact effective object inside multiquark systems.
Meanwhile, the spin triplet flavor symmetric diquark is named as ``bad diquarks", whose color configuration is still $\bar 3$~\cite{Wilczek:2004im}. Within such a configuration the quark-quark interaction becomes repulsive and the diquark structure may not be stable.

The simplest color singlet system containing the good light diquark is the  antitriplet singly heavy baryons~\cite{Kim:2024tbf}.  
As shown by the total wave functions (see App.~\ref{app:wavefunction}), the light quarks of $\Lambda_c$ and $\Xi_c$ form the scalar diquarks, while the sextet single-charm baryons contain the spin triplet diquark. In the light quark sector the $u$-$d$ quarks of the $\Lambda$ and $\Sigma^0$ are isosinglet and isotriplet, respectively.
These differences account for the mass difference between $\Lambda$ and $\Sigma^0$. Since the spin-dependent interactions between $s$ quark and $ud$ quarks are significant~\cite{Wilczek:2004im}, the light baryons are not the ideal place for studying the light diquark directly. 

The heavy-light diquark is also an effective object that is widely discussed in the literature of exotic states containing the heavy quark, especially the hidden charm/bottom tetraquarks. For a review of heavy-light diquarks, one can refer to Ref.~\cite{Shifman:2024kfj} and references therein. Interestingly, it is concluded by Ref.~\cite{Shifman:2024kfj} that the current single bottom baryons do not support the existence of heavy-light diquarks. For the single charm baryons, the diquark of light quarks is usually concerned~\cite{Selem:2006nd}. Although the traditional charmed baryons may not be the ideal place to study the heavy-light diquark since that $c$ quark is much lighter than $b$ quark~\cite{Shifman:2024kfj}, we are interested in how the spatial distribution of the heavy-light quark reflects their correlations in terms of the diquark structure.

For singly heavy baryons, it is not clear whether the heavy baryons behave like a three-quark system or with compact diquark degrees of freedom~\cite{Garcia-Tecocoatzi:2022zrf}. To be more specific, given that  the $[ud]$ diquark within $\Lambda_c$ be the ideal candidate for the ``good diquark", should it behave more likely to be a compact structure or still have spatial distribution? Intuitively, these two scenarios will have significant impact on the spectrum and decay properties.  
The decay behaviors of the single charm baryons may provide useful information about the property of diquarks. 
Some efforts can be found in the literature~\cite{Niu:2020gjw, Ortiz-Pacheco:2024qcf,Zhou:2023wrf,Yu:2022ymb,Garcia-Tecocoatzi:2022zrf,Hernandez:2011tx,Zhong:2007gp,Nagahiro:2016nsx,Lu:2018utx,Lu:2019rtg,Arifi:2021orx}.

As the lightest singly charmed baryons, $\Lambda_c$ decays only via the weak interaction. Here, we concentrate on the hadronic weak decay, which involves both the weak and strong interactions. In this work, two decay modes, $\Lambda_c\to \Lambda K^+$ and $\Lambda_c\to \Sigma^0 K^+$, are investigated in the framework of the non-relativistic constituent quark model (NRCQM) ~\cite{Close:1979bt,LeYaouanc:1988fx}. An interesting difference between these two decay modes is that the constituent quarks of the final baryons are the same but with different isospins carried by the light quarks. This discrepancy provides a good opportunity to investigate the light diquark of $\Lambda_c$, which is a ``good diquark", and its evolution to the final states, where both ``good" and ``bad diquark" are present.

Within the framework of NRCQM, all the hadronic weak decay mechanisms, which include the direct meson emission (DME) process, color suppressed (CS) process and pole terms~\cite{LeYaouanc:1988fx}, can be  described self-consistently~\cite{Niu:2020gjw}. In addition, the spatial distributions of the $[ud]$  and $[cq]$ diquarks in the charmed baryons can be described by parameters of the spatial wave functions. Thus, the effects of spatial distribution of diquarks can be conveniently investigated via the response of observables, e.g. the decay branching ratio, to the diquark parameters.

This paper is organized as follows. A brief introduction of NRQCM is presented in Sec.~\ref{sec:framework}. The results and discussions are given in Sec.~\ref{sec:result}. The last section does not discussed in the summary. The $\mathrm{SU}(3)\otimes\mathrm{O}(3)$ wave functions are listed in the Appendix.

\section{Framework}
\label{sec:framework}
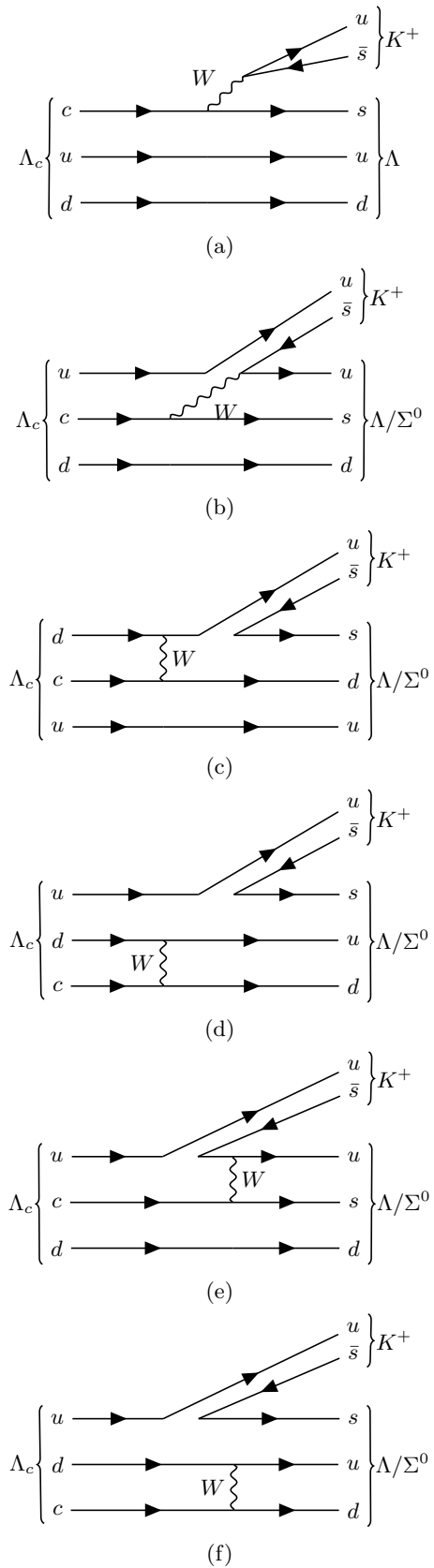
\begin{figure}[htbp]
\centering
\begin{subfigure}[h ]{0.45\textwidth}
\begin{tikzpicture}[line width=0.6pt]
\begin{feynman}
\vertex (a1) {$ c $};
\vertex[right=2.5cm of a1] (w1);
\vertex[right=2cm of a1] (a2);
\vertex[right=2cm of a2] (a3){$ s $};
\vertex[below=2em of a1] (b1) {$ u $};
\vertex[below=2em of a2] (b2) ;
\vertex[below=2em of a3] (b3) {$ u $};
\vertex[below=2em of b1] (c1) {$ d $};
\vertex[below=2em of b2] (c2);
\vertex[below=2em of b3] (c3) {$ d $};
\vertex[above=4em of a3]   (d2) {$u$};
\vertex[above=2.5em of a3] (d3) {$\bar s$};
\vertex[above=1.5em of w1] (e1);
\diagram* {
(a1) -- [fermion] (a2) -- [fermion] (a3),
(b1) -- [fermion] (b2) -- [fermion] (b3),
(c1) -- [fermion] (c2) -- [fermion] (c3),
(e1) -- [fermion] (d2),
(d3) -- [fermion] (e1),
(a2) -- [boson, edge label=$W$] (e1),
};
\draw [decoration={brace}, decorate] (c1.south west) -- (a1.north west)
node [pos=0.5, left] {$\Lambda_c$};
\draw [decoration={brace}, decorate] (a3.north east) -- (c3.south east)
node [pos=0.5, right] {$\Lambda$};
\draw [decoration={brace}, decorate] (d2.north east) -- (d3.south east)
node [pos=0.5, right] {$K^+$};
\end{feynman}
\end{tikzpicture}
\caption{ }
\label{fig:dpe}
\end{subfigure}
~
\begin{subfigure}[htbp! ]{0.45\textwidth}
\begin{tikzpicture}[line width=0.6pt]
\begin{feynman}
\vertex (a1) {$ u $};
\vertex[right=2cm of a1] (a2);
\vertex[right=4cm of a1] (a3){$ u $};
\vertex[below=2em of a1] (b1) {$ c $};
\vertex[right=1.5 of b1] (b2) ;
\vertex[below=2em of a3] (b3) {$ s $};
\vertex[below=2em of b1] (c1) {$ d $};
\vertex[right=1.5cm of c1] (c2);
\vertex[below=2em of b3] (c3) {$ d $};
\vertex[above=4 em of a3]  (d1) {$u$};
\vertex[above=2.8em of a3] (d2) {$\bar s$};
\vertex[right=0.5cm of a2] (w1);
\vertex[right=1.5 of b1] (w2) ;
\diagram* {
(a1) -- [fermion] (a2),
(w1) -- [fermion] (a3),
(b1) -- [fermion] (w2) -- [fermion] (b3),
(c1) -- [fermion] (c2) -- [fermion] (c3),
(a2) -- [fermion] (d1),
(d2) -- [fermion] (w1),
(w1) -- [boson, edge label=$W$] (w2),
};
\draw [decoration={brace}, decorate] (c1.south west) -- (a1.north west)
node [pos=0.5, left] {$\Lambda_c$};
\draw [decoration={brace}, decorate] (a3.north east) -- (c3.south east)
node [pos=0.5, right] {$\Lambda/\Sigma^0$};
\draw [decoration={brace}, decorate] (d1.north east) -- (d2.south east)
node [pos=0.5, right] {$K^+$};
\end{feynman}
\end{tikzpicture}
\caption{}
\end{subfigure}\\

\begin{subfigure}[htbp! ]{0.45\textwidth}
\begin{tikzpicture}[line width=0.6pt]
\begin{feynman}
\vertex (a1) {$ d $};
\vertex[right=1.5cm of a1] (a2);
\vertex[right=0.5cm of a2] (a3);
\vertex[right=0.5cm of a3] (a4);
\vertex[right=1.5cm of a4] (a5){$ s $};
\vertex[below=2em of a1] (b1) {$ c $};
\vertex[below=2em of a2] (b2) ;
\vertex[below=2em of a5] (b3) {$ d $};
\vertex[below=2em of b1] (c1) {$ u $};
\vertex[below=2em of b2] (c2);
\vertex[below=2em of b3] (c3) {$ u $};
\vertex[above=4em of a5]   (d1) {$u$};
\vertex[above=2.8em of a5] (d2) {$\bar s$};
\diagram* {
(a1) -- [fermion] (a3),
(a4) -- [fermion] (a5),
(b1) -- [fermion] (b2) -- [fermion] (b3),
(c1) -- [fermion] (c2) -- [fermion] (c3),
(a3) -- [fermion] (d1),
(d2) -- [fermion] (a4),
(a2) -- [boson, edge label=$W$] (b2),
};
\draw [decoration={brace}, decorate] (c1.south west) -- (a1.north west)
node [pos=0.5, left] {$\Lambda_c$};
\draw [decoration={brace}, decorate] (a5.north east) -- (c3.south east)
node [pos=0.5, right] {$\Lambda/\Sigma^0$};
\draw [decoration={brace}, decorate] (d1.north east) -- (d2.south east)
node [pos=0.5, right] {$K^+$};
\end{feynman}
\end{tikzpicture}
\caption{}
\end{subfigure} 
~
\begin{subfigure}[htbp! ]{0.45\textwidth}
\begin{tikzpicture}[line width=0.6pt]
\begin{feynman}
\vertex (a1) {$ u $};
\vertex[right=1.5cm of a1] (a2);
\vertex[right=0.5cm of a2] (a3);
\vertex[right=0.5cm of a3] (a4);
\vertex[right=1.5cm of a4] (a5){$ s $};
\vertex[below=2em of a1] (b1) {$ d $};
\vertex[below=2em of a2] (b2) ;
\vertex[below=2em of a5] (b3) {$ u $};
\vertex[below=2em of b1] (c1) {$ c $};
\vertex[below=2em of b2] (c2);
\vertex[below=2em of b3] (c3) {$ d $};
\vertex[above=4em of a5]   (d1) {$u$};
\vertex[above=2.8em of a5] (d2) {$\bar s$};
\diagram* {
(a1) -- [fermion] (a3),
(a4) -- [fermion] (a5),
(b1) -- [fermion] (b2) -- [fermion] (b3),
(c1) -- [fermion] (c2) -- [fermion] (c3),
(a3) -- [fermion] (d1),
(d2) -- [fermion] (a4),
(c2) -- [boson, edge label=$W$] (b2),
};
\draw [decoration={brace}, decorate] (c1.south west) -- (a1.north west)
node [pos=0.5, left] {$\Lambda_c$};
\draw [decoration={brace}, decorate] (a5.north east) -- (c3.south east)
node [pos=0.5, right] {$\Lambda/\Sigma^0$};
\draw [decoration={brace}, decorate] (d1.north east) -- (d2.south east)
node [pos=0.5, right] {$K^+$};
\end{feynman}
\end{tikzpicture}
\caption{}
\end{subfigure} 
\\
\begin{subfigure}[htbp! ]{0.45\textwidth}
\begin{tikzpicture}[line width=0.6pt]
\begin{feynman}
\vertex (a1) {$ u $};
\vertex[right=1.5cm of a1] (a2);
\vertex[right=0.5cm of a2] (a3);
\vertex[right=0.5cm of a3] (a4);
\vertex[right=1.5cm of a4] (a5){$ u $};
\vertex[below=2em of a1] (b1) {$ c $};
\vertex[below=2em of a4] (b2) ;
\vertex[below=2em of a5] (b3) {$ s $};
\vertex[below=2em of b1] (c1) {$ d $};
\vertex[below=2em of b2] (c2);
\vertex[below=2em of b3] (c3) {$ d $};
\vertex[above=4em of a5]   (d1) {$u$};
\vertex[above=2.9em of a5] (d2) {$\bar s$};
\vertex[right=2.5cm of a1] (w1);
\vertex[below=2em of w1] (w2);
\diagram* {
(a1) -- [fermion] (a2),
(a3) -- [fermion] (a5),
(b1) -- [fermion] (b2) -- [fermion] (b3),
(c1) -- [fermion] (c2) -- [fermion] (c3),
(a2) -- [fermion] (d1),
(d2) -- [fermion] (a3),
(w1) -- [boson, edge label=$W$] (w2),
};
\draw [decoration={brace}, decorate] (c1.south west) -- (a1.north west)
node [pos=0.5, left] {$\Lambda_c$};
\draw [decoration={brace}, decorate] (a5.north east) -- (c3.south east)
node [pos=0.5, right] {$\Lambda/\Sigma^0$};
\draw [decoration={brace}, decorate] (d1.north east) -- (d2.south east)
node [pos=0.5, right] {$K^+$};
\end{feynman}
\end{tikzpicture}
\caption{}
\end{subfigure}
~
\begin{subfigure}[htbp! ]{0.45\textwidth}
\begin{tikzpicture}[line width=0.6pt]
\begin{feynman}
\vertex (a1) {$ u $};
\vertex[right=1.5cm of a1] (a2);
\vertex[right=0.5cm of a2] (a3);
\vertex[right=0.5cm of a3] (a4);
\vertex[right=1.5cm of a4] (a5){$ s $};
\vertex[below=2em of a1] (b1) {$ d $};
\vertex[below=2em of a4] (b2) ;
\vertex[below=2em of a5] (b3) {$ u $};
\vertex[below=2em of b1] (c1) {$ c $};
\vertex[below=2em of b2] (c2);
\vertex[below=2em of b3] (c3) {$ d $};
\vertex[above=4em of a5]   (d1) {$u$};
\vertex[above=2.9em of a5] (d2) {$\bar s$};
\vertex[right=2.5cm of a1] (w1);
\vertex[below=2em of w1] (w2);
\diagram* {
(a1) -- [fermion] (a2),
(a3) -- [fermion] (a5),
(b1) -- [fermion] (b2) -- [fermion] (b3),
(c1) -- [fermion] (c2) -- [fermion] (c3),
(a2) -- [fermion] (d1),
(d2) -- [fermion] (a3),
(c2) -- [boson, edge label=$W$] (w2),
};
\draw [decoration={brace}, decorate] (c1.south west) -- (a1.north west)
node [pos=0.5, left] {$\Lambda_c$};
\draw [decoration={brace}, decorate] (a5.north east) -- (c3.south east)
node [pos=0.5, right] {$\Lambda/\Sigma^0$};
\draw [decoration={brace}, decorate] (d1.north east) -- (d2.south east)
node [pos=0.5, right] {$K^+$};
\end{feynman}
\end{tikzpicture}
\caption{}
\end{subfigure}
\caption{The transition diagrams of the $\Lambda_c\to  \Lambda/\Sigma^0~ K^+$ process. (a) DME process; (b) CS process; (c)-(f) pole terms. }
\label{fig:feynmann}
\end{figure} 

The transition diagrams for the $\Lambda_c \to \Lambda K^+$ and $\Lambda_c \to \Sigma^0 K^+$ processes are given in Fig.~\ref{fig:feynmann}.
The subfigure (a) is the DME process, where the quark pair created from the emitted W boson is hadronized into one final-state hadron. The subfigure (b) is the CS process, where the quark pair created from the emitted W boson is hadronized into different final hadrons. 
The subfigure (c), (d) (e), (f) are the pole terms diagrams. Those four processes are categorized into two cases. One is that the $W$ boson is exchanged between two quarks which are not involved in the emitted meson, i.e. (d) and (f), denoted as WS. That means the strong (S) and weak (W) interactions happen in order. The other one, denoted as SW ((c) and (e)), is that the $W$ boson is exchanged between quarks, and the two created quarks split and are combined to two different final-state hadrons. 
All transition processes are calculated in the NRCQM framework. The standard $\mathrm {SU(6)} \otimes \mathrm{O}(3)$ hadron wave functions~\cite{LeYaouanc:1988fx} listed in App.~\ref{app:wavefunction} are employed. The relevant operators, both weak interaction and strong interaction,  are shortly introduced in the following.

\subsection{The non-relativistic form of the weak interaction}

There are two kinds of weak transition processes contribute to the SW processes for this two channels. The first one is $cs\to s u$ and the second one is $cd \to du$. For the DME and CS processes,  there is only one kind of operator $c\to s (u \bar s)$ that contributes to the weak transition vertex.
In the NRCQM frame work, the operators that induce these weak transition processes can be obtained with the four-fermion interaction ~\cite{LeYaouanc:1978ef, LeYaouanc:1988fx} and the effective Hamiltonian can be expressed as:
\begin{align}
H_W=\frac{G_F}{\sqrt 2}\int d \bm x \frac12 \{J^{-,\mu}(\bm x),J^{+}_{\mu}(\bm x) \},
\end{align}
where 
\begin{align}
J^{+,\mu}(\bm x)&=
\begin{pmatrix}\bar u&\bar c \end{pmatrix}
\gamma^\mu(1-\gamma_5)
\begin{pmatrix}\cos \theta_C & \sin \theta_C \\ -\sin \theta_C &\cos \theta_C \end{pmatrix} \begin{pmatrix} d\\s \end{pmatrix}, 
\end{align}
\begin{align}
J^{-,\mu}(\bm x)&=
\begin{pmatrix}\bar d &\bar s \end{pmatrix} \begin{pmatrix}\cos \theta_C & -\sin \theta_C \\ \sin \theta_C &\cos \theta_C \end{pmatrix} \gamma^\mu(1-\gamma_5)
\begin{pmatrix} u\\c \end{pmatrix}.
\end{align}
$\theta_C$ is the Cabibbo angle. The weak transition operators that contribute to $cs\to s u$, $cd \to du$ and $c\to s (u \bar s)$ are written as
\begin{widetext}
\begin{align}
H_{W,cs\to su}&=\frac{G_F}{\sqrt{2}} V_{u s} V_{c s} \frac{1}{(2 \pi)^3} \delta^3\left(\boldsymbol{p}_c+\boldsymbol{p}_s-\boldsymbol{p}_u-\boldsymbol{p}_s\right) \bar{u}\left(\boldsymbol{p}_s\right) \gamma_\mu\left(1-\gamma_5\right) u\left(\boldsymbol{p}_c\right) \bar{u}\left(\boldsymbol{p}_u\right) \gamma^\mu\left(1-\gamma_5\right) u\left(\boldsymbol{p}_s\right)\hat{\alpha} \hat{\beta}, \\
H_{W,cd\to du}&=\frac{G_F}{\sqrt{2}} V_{u d} V_{c d} \frac{1}{(2 \pi)^3} \delta^3\left(\boldsymbol{p}_c+\boldsymbol{p}_d-\boldsymbol{p}_u-\boldsymbol{p}_d\right) \bar{u}\left(\boldsymbol{p}_d\right) \gamma_\mu\left(1-\gamma_5\right) u\left(\boldsymbol{p}_c\right) \bar{u}\left(\boldsymbol{p}_u\right) \gamma^\mu\left(1-\gamma_5\right) u\left(\boldsymbol{p}_d\right)\hat{\alpha}' \hat{\beta}', \\
H_{W,c\to s \bar s u}&=\frac{G_F}{\sqrt{2}} V_{u s} V_{c s} \frac{1}{(2 \pi)^3} \delta^3\left(\boldsymbol{p}_c-\boldsymbol{p}_s-\boldsymbol{p}_u-\boldsymbol{p}_{\bar s}\right) \bar{u}\left(\boldsymbol{p}_s\right) \gamma_\mu\left(1-\gamma_5\right) u\left(\boldsymbol{p}_c\right) \bar{u}\left(\boldsymbol{p}_u\right) \gamma^\mu\left(1-\gamma_5\right) v\left(\boldsymbol{p}_{\bar s}\right) \hat{\alpha}_1 \hat{I}^{\prime}_K.
\end{align}
\end{widetext}
where $V_{cs}$, $V_{us}$, $V_{cd}$ and $V_{ud}$ are the Cabibbo-Kobayashi-Maskawa (CKM) matrix elements. $\hat I'_K=b^\dagger_u b_u$ is the isospin operator for the $K^+$ production for the color suppressed process. $b^\dagger_{u}$ and $b_{u}$ are the creation and annihilation operators of the $u$ quark. $\hat\alpha^{(\prime)}$ and $\hat\beta^{(\prime)}$ are the flavor-changing operators with $\hat\alpha c=s,~\hat\beta s=u$ and $\hat\alpha^\prime c=d,~\hat\beta^\prime d=u$. In the non-relativistic limit, the transition operator $H_{W}$ for the $cs\to su$ process is rewritten as:
\begin{widetext}
\begin{align}
\label{eq:HW22}
H_{W,cs\to s u}^{\mathrm{P C}}&= \frac{G_F}{\sqrt{2}} V_{u s} V_{c s} \frac{1}{(2 \pi)^3} \sum_{i \neq j} \hat{\alpha}_i \hat{\beta} \delta^3\left(\boldsymbol{p}_i^{\prime}+\boldsymbol{p}_j^{\prime}-\boldsymbol{p}_i-\boldsymbol{p}_j\right)\left(1-\left\langle s_{z, i}^{\prime}\left|\boldsymbol{\sigma}_{\boldsymbol{i}}\right| s_{z, i}\right\rangle\left\langle s_{z, j}^{\prime}\left|\boldsymbol{\sigma}_j\right| s_{z, j}\right\rangle\right), \\
H_{W, cs\to s u}^{\mathrm{P V}}&= \frac{G_F}{\sqrt{2}}V_{u s} V_{c s} \frac{1}{(2 \pi)^3} \sum_{i \neq j} \hat{\alpha}_i \hat{\beta} \delta^3\left(\boldsymbol{p}_i^{\prime}+\boldsymbol{p}_j^{\prime}-\boldsymbol{p}_i-\boldsymbol{p}_j\right)  \notag  \\
&\times\left\{-\left(\left\langle s_{z, i}^{\prime}\left|\boldsymbol{\sigma}_{\boldsymbol{i}}\right| s_{z, i}\right\rangle-\left\langle s_{z, i}^{\prime}\left|\boldsymbol{\sigma}_j\right| s_{z, j}\right\rangle\right)\left[\left(\frac{\boldsymbol{p}_i}{2 m_i}-\frac{\boldsymbol{p}_j}{2 m_j}\right)+\left(\frac{\boldsymbol{p}_i^{\prime}}{2 m_i^{\prime}}-\frac{\boldsymbol{p}_j^{\prime}}{2 m_j^{\prime}}\right)\right]\right. \notag \\
& \left.+i\left(\left\langle s_{z, i}^{\prime}\left|\boldsymbol{\sigma}_i\right| s_{z, i}\right\rangle \times\left\langle s_{z, i}^{\prime}\left|\boldsymbol{\sigma}_j\right| s_{z, j}\right\rangle\right)\left[\left(\frac{\boldsymbol{p}_i}{2 m_i}-\frac{\boldsymbol{p}_j}{2 m_j}\right)-\left(\frac{\boldsymbol{p}_i^{\prime}}{2 m_i^{\prime}}-\frac{\boldsymbol{p}_j^{\prime}}{2 m_j^{\prime}}\right)\right]\right\},
\end{align}
\end{widetext}
where the superscript PC (PV) stands for the parity conserved (violated) operator $H_{W,cd\to s u}^{\mathrm{P C (PV)}}$. The subscripts $i$ and $j$ are used to indicate which quark acts in the operators. Similarly, we can obtain the $H_{W,cd\to d u}^{\mathrm{P C/PV}}$ operator.

\begin{figure}[htbp!]
\begin{center}
\begin{tikzpicture}[line width=0.6pt]
\begin{feynman}
\vertex (a1) {$ 1 $};
\vertex[right=2cm of a1] (a2);
\vertex[right=0.5cm of a2](a4);
\vertex[right=4cm of a1] (a3){$ 5 $};
\vertex[below=2em of a1] (b1) {$ 3 $};
\vertex[right=1.5 of b1] (b2) ;
\vertex[below=2em of a3] (b3) {$ 3' $};
\vertex[below=2em of b1] (c1) {$ 2 $};
\vertex[right=1.5cm of c1] (c2);
\vertex[below=2em of b3] (c3) {$ 2 $};
\vertex[above=4 em of a3]  (d1) {$1$};
\vertex[above=2.8em of a3] (d2) {$4$};
\vertex[right=0.5cm of a2] (w1);
\vertex[right=1.5 of b1] (w2) ;
\diagram* {
(a1) -- [fermion] (a2),
(a4) -- [fermion] (a3),
(b1) -- [fermion] (w2) -- [fermion] (b3),
(c1) -- [fermion] (c2) -- [fermion] (c3),
(a2) -- [fermion] (d1),
(d2) -- [fermion] (w1),
(w1) -- [boson, edge label=$W$] (w2),
};
\end{feynman}
\end{tikzpicture}
\end{center}
\caption{The color suppressed transitions at the quark level. The charm quark is always labeled as the third particle.}
\label{fig:cs}
\end{figure}
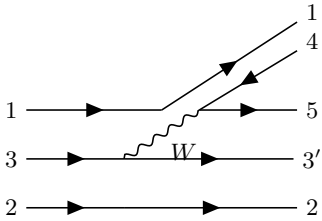

In the non-relativistic limit, the operators for the $c \to s (u\bar s) $ process are written as:
\begin{widetext}
\begin{align}
\label{eq:HW13}
H_{W,c \to s \bar s u}^{\mathrm{P C}}&=\frac{G_F}{\sqrt{2}} V_{u s} V_{c s} \frac{\beta}{(2 \pi)^3} \delta^3\left(\boldsymbol{p}_3-\boldsymbol{p}_3^{\prime}-\boldsymbol{p}_4-\boldsymbol{p}_5\right)\left\{\left\langle s_3^{\prime}|I| s_3\right\rangle\left\langle s_5 \bar{s}_4|\boldsymbol{\sigma}| 0\right\rangle\left(\frac{\boldsymbol{p}_5}{2 m_5}+\frac{\boldsymbol{p}_4}{2 m_4}\right)\right.  \notag  \\
& -\left[\left(\frac{\boldsymbol{p}_3^{\prime}}{2 m_3{ }^{\prime}}+\frac{\boldsymbol{p}_3}{2 m_3}\right)\left\langle s_3^{\prime}|I| s_3\right\rangle-i\left\langle s_3^{\prime}|\boldsymbol{\sigma}| s_3\right\rangle \times\left(\frac{\boldsymbol{p}_3}{2 m_3}-\frac{\boldsymbol{p}_3^{\prime}}{2 m_3^{\prime}}\right)\right]\left\langle s_5 \bar{s}_4|\boldsymbol{\sigma}| 0\right\rangle  \notag  \\
& -\left\langle s_3^{\prime}|\boldsymbol{\sigma}| s_3\right\rangle\left[\left(\frac{\boldsymbol{p}_5}{2 m_5}+\frac{\boldsymbol{p}_4}{2 m_4}\right)\left\langle s_5 \bar{s}_4|I| 0\right\rangle-i\left\langle s_5 \bar{s}_4|\boldsymbol{\sigma}| 0\right\rangle \times\left(\frac{\boldsymbol{p}_4}{2 m_4}-\frac{\boldsymbol{p}_5}{2 m_5}\right)\right]   \notag \\
& \left.+\left\langle s_3^{\prime}|\boldsymbol{\sigma}| s_3\right\rangle\left(\frac{\boldsymbol{p}_3^{\prime}}{2 m_3^{\prime}}+\frac{\boldsymbol{p}_3}{2 m_3}\right)\left\langle s_5 \bar{s}_4|I| 0\right\rangle\right\} \hat{\alpha}_3 \hat{I}_K^{\prime},   
\end{align}
\end{widetext}
\begin{widetext}
\begin{align}
H_{W,c \to s \bar s u}^{\mathrm{P V}}&= \frac{G_F}{\sqrt{2}} V_{u s} V_{c s} \frac{\beta}{(2 \pi)^3} \delta^3\left(\boldsymbol{p}_3-\boldsymbol{p}_3^{\prime}-\boldsymbol{p}_4-\boldsymbol{p}_5\right)\left(-\left\langle s_3^{\prime}|I| s_3\right\rangle\left\langle s_5 \bar{s}_4|I| 0\right\rangle+\left\langle s_3^{\prime}|\boldsymbol{\sigma}| s_3\right\rangle\left\langle s_5 \bar{s}_4|\boldsymbol{\sigma}| 0\right\rangle\right) \hat{\alpha}_3 \hat{I}_K^{\prime} \ ,
\end{align}
\end{widetext}
where $I$ is the two-dimension identity matrix.
We have specify the quark which the operator acts on and the quark labels are illustrated by Fig.~\ref{fig:cs}. $\beta$ is a symmetry factor which equals to $1$ for the DME process and $2/3$ for the CS process.
The $4$th particle is an anti-quark and its spin is labeled with $\bar s_4$. In order to evaluate the spin matrix element such as $\left\langle s_5 \bar{s}_4|I| 0\right\rangle$, the particle-hole conjugation~\cite{Racah:1942gsc} is employed:
\begin{align}
\langle j,-m| \to(-1)^{j+m}|j,m\rangle.
\end{align}

\subsection{The non-relativistic form of the quark-meson interaction}

As for the strong interaction between the light meson and constituent quarks, it can be described with the chiral quark model~\cite{Manohar:1983md, Zhao:2002id, Zhong:2007gp}. The Hamiltonian at leading order~\cite{Manohar:1983md, Zhong:2007gp} reads as:
\begin{align}
\label{equ:Hpi}
H_{\chi}=\int d \bm x \frac{g_A^q}{f_m}\bar q(\bm x) \gamma_\mu \gamma_5 q(\bm x) \partial^\mu \phi(\bm x) ,
\end{align}
where $q(\bm x)$ is the light quark field. $g_A^q$ is the quark axial vector coupling constant. Here we can adopt that $g_A^q=1$~\cite{Arifi:2021orx}. $f_m$ is the pseudoscalar meson decay constant depending on which meson are involved in the interaction. The decay constants of $K$ is $f_K=1.2 f_\pi$ with $f_\pi=94$ MeV. $\phi$ is the pseudoscalar octet and expressed as 
\begin{align}
\phi=\left(\begin{array}{ccc}
\frac{1}{\sqrt 2}\pi^0+\frac{1}{\sqrt 6}\eta & \pi^+ & K^+ \\
\pi^- & -\frac{1}{\sqrt 2}\pi^0+\frac{1}{\sqrt 6}\eta  & K^0\\
K^-   &\bar K^0  & -\sqrt{\frac{2}{3}}\eta
\end{array}\right).
\end{align}

The non-relativistic form of $H_\chi$ reads as:
\begin{widetext}
\begin{align}
H_{\chi}=\frac{1}{\sqrt{(2\pi)^3 2\omega_m}} \sum_{j=1}^2 \frac{1}{f_m}\left[\omega_m\left(\frac{\bm\sigma\cdot \bm p^j_f}{2m_f}+\frac{\bm\sigma\cdot \bm p^j_i}{2m_i}\right) -\bm \sigma \cdot \bm k\right] \hat I^j_\chi \delta^3(\bm p^j_f+\bm k-\bm p_i^j).
\end{align}
$\omega_m$ and $\bm k$ are the energy and momentum of the pseudoscalar meson, respectively. $\bm p^j_i$ and $\bm p^j_f$ are the initial and final momentum of the $j$th light quark, respectively. $\hat I^j_\chi$ is the isospin operator acting on the $j$th light quark and is written as:
\begin{align}
\hat I^j_\chi=\begin{cases}
b^\dagger_s b_u,                      &\mathrm{for}~ K^+,\\
b^\dagger_u b_s,                      &\mathrm{for}~ K^-.
\end{cases}
\end{align}
\end{widetext}

\subsection{Amplitudes, asymmetry parameter and decay width}
\begin{widetext}
The amplitudes of the DME, CS, WS and SW processes are constructed as follows:
\begin{align}
&\mathcal M_{\mathrm{CS},\mathrm{PC/PV}}^{J_f,J_f^z;J_i,J_i^z}=\langle B_f(\bm P_f;J_f,J_f^z);M(\bm k)|H^{\mathrm{PC/PV}}_{W,c\to s \bar s u}|B_c(\bm P_i;J_i,J_i^z) \rangle,\\
&\mathcal M_{\mathrm{DME},\mathrm{PC/PV}}^{J_f,J_f^z;J_i,J_i^z}=\langle B_f(\bm P_f;J_f,J_f^z);M(\bm k)|H^\mathrm{PC/PV}_{W,c\to s \bar s u}|B_c(\bm P_i;J_i,J_i^z) \rangle, \\
&\mathcal M_{\mathrm{WS};\mathrm{PC/PV}}^{J_f,J_f^z;J_i,J_i^z}\notag  \\
&=\langle B_f(\bm P_f;J_f,J_f^z);M(\bm k) | H_\chi | B_m(\bm P_i;J_i,J_i^z) \rangle \frac{i}{ \slashed p_{B_m}-m_{B_m} +i \frac{\Gamma_{B_m}}{2} } \langle B_m(\bm P_i;J_i,J_i^z) | H^{\mathrm{PC/PV}}_{W,cd \to su} | B_c(\bm P_i;J_i,J_i^z) \rangle,\\
&\mathcal M_{\mathrm{SW};\mathrm{PC/PV}}^{J_f,J_f^z;J_i,J_i^z}\notag  \\
&= \langle B_f(\bm P_f;J_f,J_f^z) | H^{\mathrm{PC/PV}}_{W,cs \to su} | B'_c(\bm P_i;J_i,J_i^z) \rangle  \frac{i}{ \slashed p_{B'_c}-m_{B'_c} +i \frac{\Gamma_{B'_c}}{2} } \langle B'_c(\bm P_i;J_i,J_i^z);M(\bm k) | H_\chi | B_c(\bm P_i;J_i,J_i^z) \rangle \notag \\
&+\langle B_f(\bm P_f;J_f,J_f^z) | H^{\mathrm{PC/PV}}_{W,cd \to du} | B'_c(\bm P_i;J_i,J_i^z) \rangle  \frac{i}{ \slashed p_{B'_c}-m_{B'_c} +i \frac{\Gamma_{B'_c}}{2} } \langle B'_c(\bm P_i;J_i,J_i^z);M(\bm k) | H_\chi | B_c(\bm P_i;J_i,J_i^z) \rangle \ ,
\end{align}
\end{widetext}
where $B_c(\bm P_i;J_i,J^z_i)$ and $B_f(\bm P_f;J_f,J^z_f)$ are the initial charmed baryon and final baryon states, respectively. $|B_m(\bm P_i;J_i,J_i^z)\rangle$ and $ |B'_c(\bm P_i;J_i,J_i^z)\rangle$ denote the intermediate light baryons and charmed baryons, respectively. In this work, only the ground states and the first excited states are considered for the PC processes and PV processes, respectively. Because the dominant contributions of the pole terms come from the processes that the intermediate states are approximately on-shell. $M(\bm k)$ is used to label the pseudoscalar meson.

With the transition amplitudes, we can obtain the decay width
\begin{align}
&\Gamma(A\to B+C)\notag \\
&=8\pi^2\frac{|\bm k|E_B E_C}{M_A}\frac{1}{2 J_A+1} \sum_\mathrm{spin} \left(|\mathcal M_{\mathrm{PC}}|^2+ |\mathcal M_{\mathrm{PV}}|^2 \right)
\end{align}
with the normalization convention 
\begin{align}
\langle B_f(P_f) |B_i(P_i) \rangle =\delta^3(\bm P_f- \bm P_i),
\end{align}
and $J_A$ is the spin of the initial state. The parity asymmetry parameter is also an important observable for the hadronic weak decay which was first introduced by Lee and Yang~\cite{Lee:1957qs}. In our work, the parity asymmetry parameter can be obtained from the parity conserved and violated amplitudes:
\begin{align}
\label{eq:ap}
\alpha=\frac{2 \operatorname{Re}\left[\mathcal M^*_\mathrm{P V} \mathcal M_\mathrm{P C}\right]}{\left|\mathcal M_\mathrm{P C}\right|^2 +\left|\mathcal M_\mathrm{P V}\right|^2 } \ ,
\end{align}
where $\mathcal M_{\mathrm{P C}}$ and $\mathcal M_{\mathrm{P V}}$ are the PC and the PV amplitudes with $J_i^z=J_f^z=1/2$, respectively.

\section{Numerical Results and discussion}
\label{sec:result}

\subsection{Parameters and inputs}
The values of quark mass used in this work are $m_u=m_d=0.3$ GeV, $m_s=0.5$ GeV and $m_c=1.8$ GeV, which are the same as those from Ref.~\cite{Niu:2025lgt}. 
According to Figs.~\ref{fig:feynmann}(c)-(f), the intermediate states of the pole terms considered in this work are as follows: 
\begin{itemize}
    \item $p,\Xi_c^0$ ,and $\Xi'^0_c$ for the PC processes,
    \item $N(1535),N(1650), |\Xi_c^0,~^2 P_\lambda\rangle, |\Xi_c^0,~^2 P_\rho\rangle, |\Xi_c^0,~^4 P_\rho\rangle,$ $|\Xi'^0_c,~^2 P_\lambda\rangle, |\Xi'^0_c,~^2 P_\rho\rangle$, and $|\Xi'^0_c,~^4 P_\lambda\rangle$  for the PV processes.
\end{itemize}
The mass of $p,\Xi_c^0,\Xi'^0_c,N(1535)$ and $N(1650)$ are taken from Particle Date Group (PDG) ~\cite{ParticleDataGroup:2024cfk}.
As for $N(1535)$ and $N(1650)$, we take two cases to describe them. The first one is that $N(1535)$ and $N(1650)$ are treated as the states of $\left [70, ^2 8 \right ]$ and $\left [70, ^4 8 \right ]$, respectively. The second one is that $N(1535)$ and $N(1650)$ are treated as the mixing states of $\left [70, ^2 8 \right ]$ and $\left [70, ^4 8 \right ]$ and is expressed as 
 \begin{align}
\left( \begin{array}{c}
N(1535) \\
N(1650)
\end{array}\right)=
\left(\begin{array}{cc}
\cos \theta & -\sin \theta \\
\sin \theta & \cos \theta 
\end{array}\right)
\left( \begin{array}{c}
\left |70, ^2 8,1,1,\frac{1}{2} \right \rangle\\
 \left |70, ^4 8,1,1,\frac{1}{2} \right \rangle
\end{array}\right).
\end{align}
$\theta$ is the mixing angle that ranges from $24^\circ$ to $32^\circ$\cite{Zhong:2011ti}.
In this work, we take $\theta=30^\circ$. The masses of $N(1535)$ and $N(1650)$ are also taken from PDG \cite{ParticleDataGroup:2024cfk} for both cases.

Due to lack of experimental data of the mass of the first excited single charmed baryons, the mass of the first excited states of $\Xi_c$ are taken from  Ref.~\cite{Garcia-Tecocoatzi:2022zrf} in a constituent quark model. In this literature, there two different models and the masses based on three-quark model are employed in our calculation. More theoretical results about the mass of the first excited single charmed baryons can be find in the Tab.~V of Ref.~\cite{Niu:2025lgt} and used as a comparison. The width of the first excited charmed baryons is also not determined in experiments. As the result, the estimated value $5\times 10^{-4}$ GeV is taken from  Ref.~\cite{Niu:2021qcc}.

\begin{table*}[htbp]
\centering
\caption{The hadron masses and widths (in unit of GeV). The `-' means the width is very small that can be neglected or unrelated to this work. The width of first excited charmed baryons are estimated values.}
\begin{ruledtabular}
\begin{tabular}{cccccccccc}
states & $p$     &$\Lambda$ &$\Sigma^0$ &$\Lambda_c$ &$K^+$   &$\Xi_c^0$ &$\Xi_c'^{0}$ 
 &$N(1535)$& $N(1650)$\\
mass   & $0.938$ &$1.116$   &$1.193$    &$2.286$     &$0.494$ &$2.47$   &$2.579$  &$\approx 1.510$ &$\approx 1.665$\\
width  &-& - & - & - & - &- &- &$0.11$ & $0.135$ \\
\hline
\multicolumn{1}{c}{\multirow{2}[0]{*}{States}}& \multicolumn{3}{l}{$\Xi^0_c$~\cite{Garcia-Tecocoatzi:2022zrf}} & \multicolumn{3}{l}{$\Xi'^0_c$~\cite{Garcia-Tecocoatzi:2022zrf}}  \\ 
&$|^2 P_\lambda\rangle$ &$|^2 P_\rho\rangle$ &$|^4 P_\rho\rangle$
&$|^2 P_\lambda\rangle$ &$|^2 P_\rho\rangle$ &$|^4 P_\lambda\rangle$ \\
mass   & $2.788$        &$ 2.935$            &$2.977$        &  $2.893$   &  $3.040$         & $2.935$\\
width  &$5\times10^{-4}$&$5\times10^{-4}$&$5\times10^{-4}$&$5\times10^{-4}$&$5\times10^{-4}$&$5\times10^{-4}$\\
\end{tabular}
\end{ruledtabular}
\label{tab:para}
\end{table*}

The values of harmonic oscillator strengths for the baryons are listed in Tab.~\ref{tab:paraAlpha}. 
Only the values of $\alpha_\rho$ are listed. Because the strong interaction is flavor independent, the strength constants $\alpha_\lambda$ and $\alpha_\rho$ are related with
\begin{align}
\label{eq:lr}
\frac{\alpha_\lambda}{\alpha_\rho}=\frac{1}{\sqrt 2} \left[\frac{3(m_1+m_2)m_3}{(m_1+m_2+m_3)m_1m_2}\right]^{1/4}.
\end{align}
More details can be find in App.~\ref{app:wavefunction}.
Considering the SU(3) flavor symmetry breaking, the value of $\alpha_\rho$ are different for the charmed and strange baryons.
Note that the value of $\alpha_\rho$ are taken empirically as in our previous work ~\cite{Niu:2020gjw,Niu:2021qcc}. This is reasonable as this work is a qualitative analysis of the diquark structures in the $\Lambda_c$.

\begin{table}[htbp]
\centering
\caption{The values of harmonic oscillator strengths (in unit of GeV). $\theta$ is the mixing angle.}
\begin{ruledtabular}
\begin{tabular}{ccc}
$\theta$   &$0^\circ$   &$30^\circ$ \\
\hline
$\alpha_\rho(S=0,C=0)$ & 0.5 &0.5 \\
$\alpha_\rho(S=1,C=0)$ &0.5  & 0.5  \\
$\alpha_\rho(S=0,C=1)$ &0.4  &0.45 \\
$\alpha_\rho(S=1,C=1)$ &0.4 &0.5 \\
\end{tabular}
\end{ruledtabular}
\label{tab:paraAlpha}
\end{table}

\subsection{The numerical results and the selection rules}

The $\Lambda$ and the $\Sigma^0$ have the same constituent quarks, so the $\Lambda_c\to\Lambda K^+$ and $\Lambda_c\to\Sigma^0 K^+$ processes share the same decay mechanisms as shown by Fig.~\ref{fig:feynmann}. While the branching ratio of the $\Lambda_c\to\Lambda K^+$ process is $(6.42\pm0.31)\%$ which is about 1.7 times as that, i.e.$(3.70\pm 0.31)\%$~\cite{ParticleDataGroup:2024cfk}, of the $\Lambda_c\to\Sigma^0 K^+$ process. 
The only apparent difference between these two channel is that the isospin of the final baryons. The $\Lambda$ ($\Sigma^0$) is an isospin singlet (triplet) state. This implies that the key to understand the different branching ratios is the configuration of $u$ and $d$ quarks.
All the values of transition amplitudes are listed in Tab.~\ref{tab:numResults}. One should note that some transition amplitudes are vanishing, as either the strong transition processes or the weak transition processes are forbidden by selection rules, such as the the ``$\Lambda$ selection rule'' making that the $N(1650)$ has no contributions to the pole terms of $\Lambda_c \to \Lambda K^+$.

Firstly, we note that the DME process has no contribution to the $\Lambda_c\to \Sigma^0K^+$. This is because that the $u$ and $d$ quarks are spectators in the DME process, as illustrated by Fig.~\ref{fig:dpe}. The isospin of the $u$ and $d$ quarks pair should not be changed. $\Lambda_c$ belongs to the SU(3) flavor $\bar 3$ representation, leaving the isospin of the $[ud]$ quarks $0$. As the $\Lambda$ ($\Sigma^0$) is an isospin singlet (triplet) state, Thus the DPE process of the $\Lambda_c\to \Lambda K^+$ ($\Lambda_c\to \Sigma^0 K^+$) process is allowed (forbidden).

Secondly, as shown in Tab.~\ref{tab:numResults}, several transition amplitudes of the pole terms are also $0$. When $\theta=0^\circ$, $N(1650)$ has no contribution to the WS process of the $\Lambda_c \to \Lambda K^+$ process,  because of the ``$\Lambda$ selection rule''~\cite{Zhao:2007xb} which leads to the vanishing transition matrix element between $N^*$ of $[ 70,^48]$ and $[ 56, ^28]$ in $N^* \to \Lambda \ K/ K^*$. 
This follows because, to keep the total wave function of $\Lambda$ completely antisymmetric, the total spin of the $[ud]$ in the $\Lambda$ has to be $S_{[ud]}=0$. The $N(1650)$ is a $[ 70,^48]$ state. So the total spin of $[ud]$ in the $N(1650)$ is $1$ as illustrated by the wave function of $N(1650)$ given in App.~\ref{app:wavefunction}. 
With the spectator approximation, the strangeness emissions in $N(1650) \to \Lambda K$, the spin of $[ud]$ must be persevered. As a consequence, this leads to a correlated vanishing transition matrix element $N(1650)\to \Lambda K^+$.

For the SW processes, $c$ quark is not involved in the strong interaction and can be treated as a spectator and the spin of $c$ is not changed. So only the light degrees of freedom need to be dealt with at the strong interaction vertex. In other words, the heavy quark spin symmetry (HQSS) is applicable for this kinds of processes. The $J^P$ of the light degrees of freedom of the charmed baryons are listed in Tab.~\ref{tab:JP}.

\begin{table}[!htbp]
\centering
\caption{The quantum number of the light degrees of freedom for charmed baryons.}
\begin{tabular}{p{2cm}l}
\hline \hline
States & $J^P$ \\
\hline
$\left |^2 S,\frac{1}{2}^+ \right \rangle_{\bar 3}$  & $0^+$  \\
$\left |^2 P_\lambda,\frac{1}{2}^- \right \rangle_{\bar 3}$  & $0^-$ or $1^-$   \\
$\left |^2 P_\rho,\frac{1}{2}^- \right \rangle_{\bar 3}$ & $1^-$  \\
$\left |^4 P_\rho,\frac{1}{2}^- \right \rangle_{\bar 3}$ & $0^-$ or $1^-$  \\
$\left |^2 S,\frac{1}{2}^+ \right \rangle_{6}$  &$0^+$  \\
$\left |^2 P_\lambda,\frac{1}{2}^- \right \rangle_{6}$ &$0^-$ or $1^-$ \\
$\left |^2 P_\rho,\frac{1}{2}^- \right \rangle_{6}$  & $1^-$ \\
$\left |^4 P_\lambda,\frac{1}{2}^- \right \rangle_{6}$ & $0^-$ or $1^-$  \\
\hline \hline
\end{tabular}
\label{tab:JP}
\end{table}

For the PC part of SW process, the strong transition processes are $B_{\bar 3} \to B_{\bar3}  K^+$ and $B_{\bar 3} \to B_6 K^+$. The first process can be presented as $0^+\to 0^+ 0^-$. In order to keep parity conservation, only odd orbital angular momentum  is allowed. However, only the $S$ wave interaction is allowed because of the total angular momentum conservation. Thus, the strong transition amplitude $B_{\bar 3} \to B_{\bar3}  K^+$ is vanishing. As for $B_{\bar 3} \to B_6 K^+$, e.g. $0^+\to 1^+ 0^-$, the $P$ wave interaction is allowed. So, as shown in Tab.~\ref{tab:numResults}, $\Xi_c^0$ has no contribution, but $\Xi'^0_c$ can contribute to the PC parts of the SW processes in $\Lambda_c\to\Lambda/\Sigma^0 K^+$.

Table~\ref{tab:numResults} also shows that the first orbital excitation state of $|\Xi_c^0, ^2P_\lambda \rangle$ and $|\Xi'^0_c,^2P_\rho \rangle$ do not contribute to the PV parts of the SW processes. Here, we also focus on the light degrees of freedom. Note that the total spin of the light quarks in $|\Xi_c^0, ^2P_\lambda \rangle$ and $|\Xi'^0_c,^2P_\rho \rangle$ is $0$. Thus, the $J^P$ of the light degrees of freedom for these two states are $1^-$. The transition of $0^+\to 1^- 0^-$ is not allowed because of the momentum conservation and parity conservation. At the quark level, one can easily find that 
\begin{align}
\langle \chi^\rho_{\frac{1}{2},-\frac{1}{2}} |\sigma_{x,y,z}^{(1)} |\langle \chi^\rho_{\frac{1}{2},-\frac{1}{2}}\rangle=0,
\end{align}
which leads to that $\Xi_c^0$, $|\Xi_c^0, ^2P_\lambda \rangle$ and $|\Xi'^0_c,^2P_\rho \rangle$ do not contribute to the strong interaction matrix elements of the SW processes.

We also find that
\begin{align}
\left \langle \Sigma^0\ K^+ \right |H_{W,cd\to du}^{\mathrm{PC}} \left |\Xi_c^{ 0}/\Xi_c'^{0} \right \rangle&=0, \notag \\
\left \langle \Sigma^0\ K^+ \right |H_{W,cd\to du}^{\mathrm{PV}} \left |\Xi_c^{ 0}/\Xi_c'^{0}, ^2 P_\lambda \right \rangle&=0.
\end{align}
while 
\begin{align}
\left \langle \Sigma^0\ K^+ \right |H_{W,cd\to du}^{\mathrm{PC}} \left |\Xi_c^{ 0}/\Xi_c'^{0} \right \rangle& \neq 0, \notag \\
\left \langle \Sigma^0\ K^+ \right |H_{W,cd\to du}^{\mathrm{PV}} \left |\Xi_c^{ 0}/\Xi_c'^{0}, ^2 P_\lambda \right \rangle& \neq 0.
\end{align}
Combined with that 
\begin{align}
\left \langle \Lambda \ K^+ \right |H_{W,cd\to du}^{\mathrm{PC}} \left |\Xi_c^{ 0}/\Xi_c'^{0} \right \rangle&\neq0, \notag \\
\left \langle \Lambda \ K^+ \right |H_{W,cd\to du}^{\mathrm{PV}} \left |\Xi_c^{ 0}/\Xi_c'^{0}, ^2P_\lambda \right \rangle&\neq0,
\end{align}
from which we can speculate that the selection rule of the weak transition processes correlate with the configuration of $[ud]$.

\begin{table*}[htbp]
\centering
\caption{The values of the $\Lambda_c\to \Lambda/\Sigma^0 K^+$ amplitudes (in unit of $10^{-9}~ \mathrm{GeV}^{-1/2}$). For the SW processes there are two kinds of weak transitions processes i.e. $cs\to su$ and $cd\to ud$ and the relevant results are labeled with $(a,b)$. Here $a$ is the amplitude for $cs\to su$ and $b$ is the amplitudes for $cd\to ud$.}
\begin{ruledtabular}
\begin{tabular}{crrllll}
\multicolumn{1}{c}{\multirow{2}[0]{*}{Parity}}         & \multicolumn{1}{c}{\multirow{2}[0]{*}{Processes}}    &    \multicolumn{1}{r}{\multirow{2}[0]{*}{States}}        &\multicolumn{2}{l}{\multirow{1}[0]{*}{$\Lambda_c\to\Lambda \ K^+$}}  &\multicolumn{2}{l}{\multirow{1}[0]{*}{$\Lambda_c \to \Sigma^0 \ K^+$ }}\\
                       &     & -         &  $\theta=0^\circ$ & $\theta=30^\circ$  &$\theta=0^\circ$ & $\theta=30^\circ$ \\
\hline
\multirow{5}[0]{*}{PC} &DPE & -          &$-6.71$ &$-6.29$&$0$  &$0$\\
                       &CS  & -          &$0.67$  &$0.63$&$-1.16$&$-1.08$\\
                       &WS  & $p$        &$-1.89$ &$-2.30$&$0.36$&$0.44$\\
                       &SW  &$\Xi_c^0$   &$(0,0)$ &$(0,0)$& $(0,0)$ &$(0,0)$\\
                       &    &$\Xi'^{0}_c$&$(0.60,1.13)$ &$(1.33,2.53)$&$(-1.14,0)$        &$(-2.45,0)$ \\
                       &Total& -         &$-6.21$ &$-4.11$&$-1.94$&$-3.10$\\
\hline
\multirow{6}[0]{*}{PV} &DPE &-           &$4.93$ &$4.48$&$0$  &$0$\\
                       &CS  &-           &$-1.10$&$-1.08$&$1.97$&$1.91$\\
                       &WS  &$N(1535)$   &$2.68-0.15i$ &$3.23-0.18i$&$0.58-0.032i$&$-0.65+0.084i$\\
                       &    &$N(1650)$   & $0$   &$0.87-0.048i$&$4.37-0.40i$&$5.24-0.47i$\\
                       &SW  &$\Xi^0_c|^2P_\rho\rangle$     &$(-0.36,0.82)$&$(-0.99,2.20)$&$(0.89,0.096)$&$(2.28,0.20)$ \\
                       &    &$\Xi^0_c|^2P_\lambda\rangle$  &$(0,0)$&$(0,0)$ &$(0,0)$&$(0,0)$\\
                       &    &$\Xi^0_c|^4P_\rho\rangle$     &$(0.55,-1.05)$&$(1.45,-2.75)$       &$(-1.05,0)$& $(-2.67,0)$  \\
                       &    &$\Xi'^0_c|^2P_\rho\rangle$    &$(0,0)$&$(0,0)$ &$(0,0)$& $(0,0)$ \\
                       &    &$\Xi'^0_c|^2P_\lambda\rangle$ &$(-0.45,-0.80)$&$(-1.19,-2.11)$&$(0.78,0.043)$&$(2.03,0.091)$  \\
                       &    &$\Xi'^0_c|^4P_\lambda\rangle$ &$(0.56,1.06)$& $(1.45,2.75)$      &$(-1.05,0)$& $(-2.66,0)$\\
                       &Total&-                            &$6.84-0.15i$& $8.32-0.23i$      &$6.63-0.43i$&$5.77-0.39i$       \\
\end{tabular}
\end{ruledtabular}
\label{tab:numResults}
\end{table*}

As shown by Tab.~\ref{tab:numResults}, the amplitudes of the non-factorizable processes could have the same order as that of the factorizable ones. It means that the non-factorizable processes may be important and need to be treated carefully considering the interference among these terms. 
Meanwhile, one should note that the contributions of $p$, $N(1535)$ and $N(1650)$ are very different for these two channels because of the ``$\Lambda$ selection rule''~\cite{Zhao:2007xb}. As a consequence, the mixing of $\left [70, ^2 8 \right ]$ and $\left [70, ^4 8 \right ]$ has noticeable impact on the total parity violating amplitudes which is reflected in the asymmetry parameters as shown by Tab.~\ref{tab:R&alpha}.

\begin{table}[htbp]
  \centering
\caption{The branching ratio ($\times 10^{-4}$) and the asymmetry parameter of $\Lambda_c \to \Lambda \ K^+$ and $\Lambda_c \to \Sigma^0 \ K^+$.}
\begin{tabular}{c|c|cc|cc}
\hline\hline
\multicolumn{1}{c|}{\multirow{2}[0]{*}{Channel}} & \multicolumn{1}{c|}{\multirow{2}[0]{*}{Br. Exp. }} & \multicolumn{2}{c|}{Br.} & \multicolumn{2}{c}{$\alpha$} \\
                            &              &$\theta=0^\circ$ &$\theta=30^\circ$    & $\theta=0^\circ$     & $\theta=30^\circ$ \\
\hline
$\Lambda_c\to\Lambda\ K^+$  &$6.42\pm0.31$ &$6.62$ &$6.69$ &$0.99$ &$0.79$ \\
$\Lambda_c\to\Sigma^0\ K^+$ &$3.70\pm0.31$ &$3.45$ &$3.09$ &$0.53$ &$0.83$\\
\hline\hline
\end{tabular}
\label{tab:R&alpha}
\end{table}

\subsection{The distribution of the diquark}

In the NRCQM framework work, the spatial distribution of the constituent quarks is described by the spatial wave function $\psi(\bm \rho,\bm \lambda)$. 
The average value of the square of displacement between quarks (i.e. kind of mean square radii) $\bm r_{ij}^2=(\bm r_i-\bm r_j)^2$ can be obtained by
\begin{align}
\langle \bm r_{ij}^2 \rangle=\langle \psi(\bm \rho, \bm \lambda)|\bm r_{ij}^2|\psi(\bm \rho, \bm \lambda)\rangle,
\end{align}
where $\psi(\bm \rho, \bm \lambda)$ is the spatial wave function in the coordinate space. $\bm \rho$ and $\bm \lambda$ are the Jacobi coordinates and the details are listed in App.~\ref{app:wavefunction}.
For the ground states, we have 
\begin{align}
\langle \bm \rho^2 \rangle=\frac{3}{2} \frac{1}{\alpha_\rho^2},~
\langle \bm \lambda^2 \rangle=\frac{3}{2} \frac{1}{\alpha_\lambda^2}.
\end{align}
Then, $\sqrt{\langle \bm \rho^2 \rangle}$, which is the average distance between the $u$ and $d$ quark, can serve as a useful quantity describing the size of the $[ud]$ diquark of the ground states of the singly charmed baryons. Thus, we can see that the larger value of $\alpha_\rho$ means that the $[ud]$ diquark is more compact. For the ground states of the singly charmed baryons, similarly, the size of the heavy-light diquark can be  defined as
\begin{align}
\langle \bm r^2_{cq} \rangle &=\langle \psi(\bm \rho, \bm \lambda)|(\bm r_3-\bm r_q)^2|\psi(\bm \rho, \bm \lambda)\rangle \notag \\
&= \langle \psi(\bm \rho, \bm \lambda)|\frac{3}{2}\bm \lambda^2+\frac{1}{2} \bm \rho^2|\psi(\bm \rho, \bm \lambda)\rangle \notag \\
&=\frac{9}{4} \frac{1}{\alpha_\lambda^2}+\frac{3}{4} \frac{1}{\alpha_\rho^2}.
\end{align}
This result indicates that the larger value of $\alpha_\rho$ or $\alpha_\lambda$, the more compact of the heavy-light diquark and that the $\alpha_\lambda$ plays a leading role.

\begin{figure}[!htbp]
\begin{center}
\includegraphics[scale=0.5]{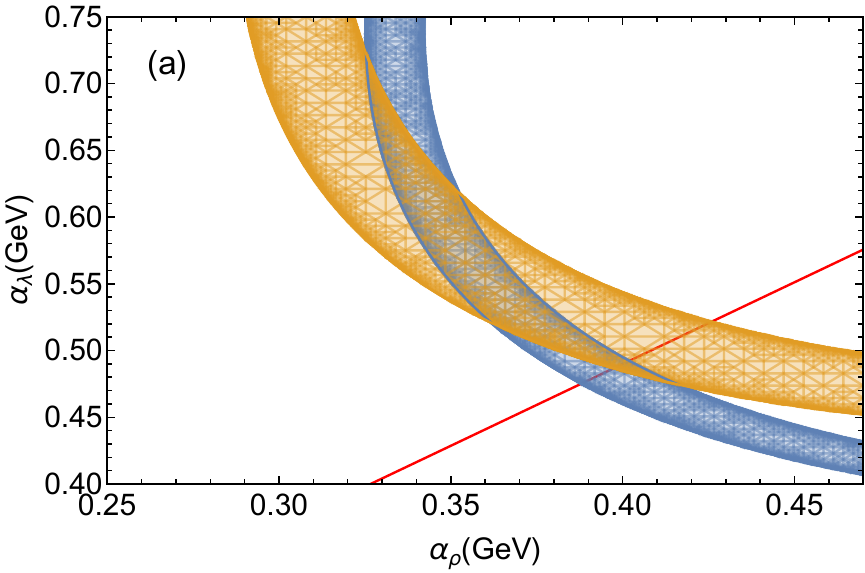}
\includegraphics[scale=0.52]{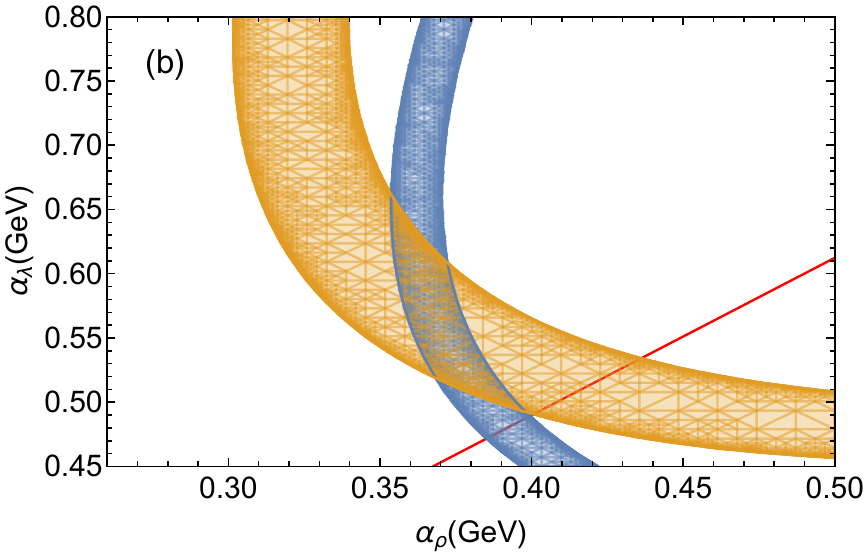}
\caption{The parameter ranges of $\alpha_\rho$ and $\alpha_\lambda$ constrained by the $\Lambda_c \to \Lambda\pi^+$ (orange band) and $\Lambda_c \to \Sigma^0 \pi^+$ (blue band) decay widths. The red solid line is determined within quark model, i.e. via Eq.~(\eqref{eq:lr}). Subfigure (a) and (b) are for the mixing angle $\theta=0^\circ$ and $\theta=30^\circ$, respectively. }
\label{fig:gamma}
\end{center}
\end{figure}

In the NRCQM framework the parameters $\alpha_\rho$ and $\alpha_\lambda$ are not independent, which are determined by the quark mass and the spring constant $K$ as shown by Eq.~\eqref{eq:arholambda}.
In this part, we assume that $\alpha_\rho$ and $\alpha_\lambda$ are independent of each other and regard them as free parameters. Then, the physical values of $\alpha_\rho$ and $\alpha_\lambda$ can be constrained by the relevant decay widths. In this work only the distribution of the good light diquark for the singly charmed baryons is concerned, ignoring the SU(3) flavor symmetry breaking effects, i.e. the same values are taken for $\alpha_{\rho/\lambda}$ for the antitriplet singly charmed baryons. Fixing the harmonic oscillator strength of the other baryons, the ranges of $\alpha_\rho$ and $\alpha_\lambda$ are constrained by the decay width as shown by Fig.~\ref{fig:gamma}. The orange and blue band are the region constrained by the decay widths of $\Lambda_c \to \Lambda\pi^+$ and $\Lambda_c \to \Sigma^0 \pi^+$, respectively. The red line is obtained with Eq.~(\eqref{eq:lr}).
Figure~\ref{fig:gamma} indicates that the allowed region of $\alpha_\rho$ and $\alpha_\lambda$ is restricted and consistent with the traditional values of NRCQM, no matter the mixing of $N(1535)$ and $N(1650)$ is considered or not. As both of these two parameters favor small values, neither the $[ud]$ quark pair nor the heavy-light diquark in the singly charmed baryons 
is point-like structure, which is consistent with the conclusion of  Ref.~\cite{Niu:2020gjw}.

We also notice that if the distribution of the $[ud]$ diquark of $\Lambda_c$ and $\Lambda$ or $\Sigma^0$ are very different from each other, the contribution of DME will be suppressed. The reason is that the overlap of the spatial wave function between the $\rho$ mode of $\Lambda_c$ and the $\rho$ mode of $\Lambda$ will be suppressed with the increase of $\alpha_\rho$ of the antitriplet charmed baryons.
This leads to that the decay width of these two channels will be saturated by the contribution of the CS and pole terms. It means that the ratio of the decay width $R_\Gamma$, which is defined as
\begin{align}
R_\Gamma=\frac{\mathrm{Br}(\Lambda_c \to \Lambda \ K^+)}{\mathrm{Br}(\Lambda_c \to \Sigma^0 \ K^+)},
\end{align}
should be close to a constant. It is determined by the flavor symmetry, with the increase of $\alpha_\rho$.

\begin{figure}[htbp!]
\begin{center}
\includegraphics[scale=0.5]{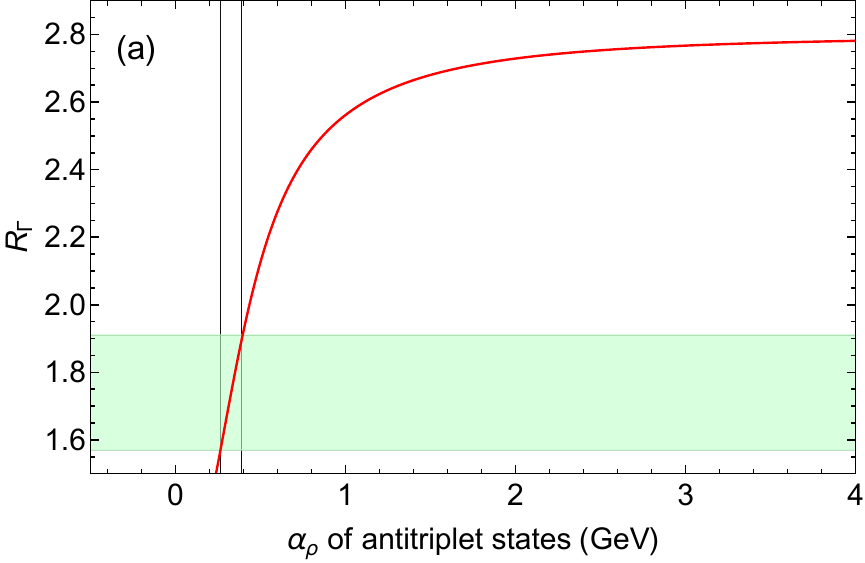}
\includegraphics[scale=0.5]{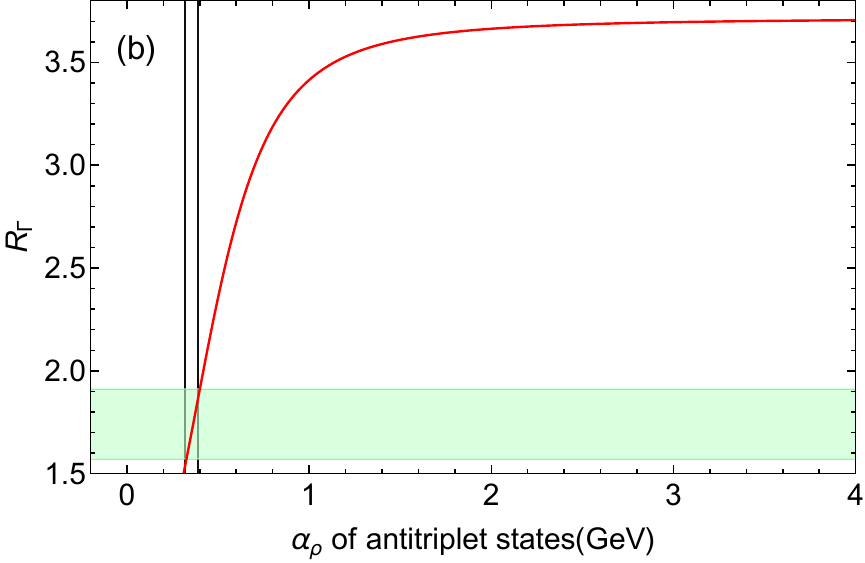}
\caption{The ratio $R_\Gamma$ varies with $\alpha_\rho$ of the antitriplet states. The parameter $\alpha_\lambda$  of the antitriplet states is fixed. The cyan band is the range of the experimental data. The two vertical lines indicate the boundary of $\alpha_\rho$ of $\Lambda_c$. The left panel (a) is for the case with $\theta=0^\circ$ and the right panel (b) is for the case with $\theta=30^\circ$.} 
\label{fig:RGamma}
\end{center}
\end{figure}

For the $cs\to s u$, $cd \to du$ or $c\to s (u \bar s)$ processes, the total isospin and the third component of the isospin of the system are all changed as: $\Delta I=1/2$ and $\Delta I_3=1/2$. Using the weak isospin analysis and ignoring the phase space difference, we can obtain that 
\begin{align}
R_\Gamma \approx  \frac{|\langle \Lambda K^+ |H_W|\Lambda_c \rangle |^2}{|\langle \Sigma^0 K^+ |H_W|\Lambda_c \rangle |^2}
=\frac{|\langle 0,0;\frac{1}{2}, \frac{1}{2}|0,0;\frac{1}{2}, \frac{1}{2} \rangle|^2}{|\langle 1,0;\frac{1}{2}, \frac{1}{2}|0,0;\frac{1}{2}, \frac{1}{2} \rangle|^2}=3.
\end{align}
Considering the SU(3) flavor symmetry breaking effects, $R_\Gamma$ should be close to 3. Figure~\ref{fig:RGamma} clearly shows that $R_\Gamma$ has an upper limit with or without considering the mixing between $\left [70, ^2 8 \right ]$ and $\left [70, ^4 8 \right ]$ with the increase of $\alpha_\rho$. Indeed, the upper limits are close to $3$ as expected.

We also note that $R_\Gamma$ is about $2.9$ with $\theta=0$ and a larger $\alpha_\rho$ value, as shown by Fig.~\ref{fig:RGamma} (a). While $R_\Gamma$ is about 3.54, as illustrated by Fig.~\ref{fig:RGamma} (a), with $\theta=30^\circ$ and $\alpha_\rho$ greater than 2 GeV.
The results depend on whether the mixing between $\left [70, ^2 8 \right ]$ and $\left [70, ^4 8 \right ]$ is considered or not.
If the mixing angle $\theta$ is set to $30^\circ$ and the other parameters are fixed, one can easily verify that the total contributions of $N(1535)$ and $N(1650)$ become larger. This leads to that the branching ratio of $\Lambda_c \to \Lambda K^+$ becomes larger, While the branching ratio of $\Lambda_c \to \Sigma^0 K^+$ becomes smaller as shown in Tab.~\ref{tab:numResults}. Thus, the $R_\Gamma$ becomes larger when the mixing between $\left [70, ^2 8 \right ]$ and $\left [70, ^4 8 \right ]$ is  considered.

\section{Summary and Outlook}

Two singly Cabibbo-suppressed hadronic weak decays of $\Lambda_c$, i.e. $\Lambda_c\to \Lambda K^+$ and $\Lambda_c\to \Sigma^0 K^+$, are investigated within the NRCQM frame work. The final baryons have different isospins, which provide an opportunity to discuss diquarks of the charmed baryons. The spatial distribution of the diquark is reflected by the harmonic oscillator parameters. The larger the harmonic oscillator parameter is, the more compact the spatial distribution is. Using the $\Lambda_c\to \Lambda K^+$ and $\Lambda_c\to \Sigma^0 K^+$ processes, we find that the allowed value region of  $\alpha_\rho$ and $\alpha_\lambda$ of the antitriplet singly charmed baryons is small and close to the commonly adopted value of the constituent quark model.
Neither the $[ud]$ diquark nor the heavy-light one turns out to favor a compact structure.

We also find that the ``$\Lambda$ selection rule'' plays an essential role for understanding why the contributions of the $N(1535)$ and the $N(1650)$ are different. Such selection rules can be useful for the search of  excited baryon states or charmed baryons in the heavy-flavor baryon hadronic weak decays.

\begin{acknowledgements}
This work is supported in part by the National Natural Science Foundation of China with Grant No.~12375073,  No.~12035007, No. 12147128, and No. 12235018. 
\end{acknowledgements}

\begin{appendix}
\section{The wave function of hadrons}
\label{app:wavefunction}
The standard $\mathrm {SU(6)} \otimes \mathrm{O}(3)$ wave functions are used in this work and the explicit baryons and pseudoscalar mesons are presented in this section. The spin-flavor wavefunctions can be expressed as $\left|N_6,^{2S+1}N_3 \right\rangle$, where $S$ stands for the total spin. $N_6 (=56, 70 \ \mathrm{or}\ 20)$ and $N_3 (=8, 10, \ \mathrm{or}\ 1)$ denote the dimension fo the SU(6) and SU(3) representation, respectively. The $\mathrm {SU(6)} \otimes \mathrm{O}(3)$ wavefunction is denoted with $\left|N_6,^{2S+1}N_3,\ N,\ L,\ J \right\rangle$ follow the convention of Refs,~\cite{Isgur:1977ef,Isgur:1978xj}. $N$ is the principal quantum number, $L$ is the total orbital angular momentum and $J$ is the total angular momentum of the hadron.

\subsection{The spin wave functions}
The spin wave functions for the three quark system are:
\begin{align}
\chi^{\rho}_{\frac{1}{2},\frac{1}{2}}&=\frac{1}{\sqrt2}\left(\uparrow \downarrow \uparrow -\downarrow \uparrow \uparrow \right) ,\\
\chi^{\lambda}_{\frac{1}{2},\frac{1}{2}}&=-\frac{1}{\sqrt6}\left( \uparrow\downarrow\uparrow+\downarrow\uparrow\uparrow-2\uparrow\uparrow\downarrow \right) , \\
\chi^{\rho}_{\frac{1}{2},-\frac{1}{2}}&=\frac{1}{\sqrt2}\left(\uparrow \downarrow \downarrow -\downarrow \uparrow \downarrow \right) ,\\
\chi^{\lambda}_{\frac{1}{2},-\frac{1}{2}}&=\frac{1}{\sqrt6}\left( \uparrow\downarrow\downarrow+\downarrow\uparrow\downarrow-2\downarrow\downarrow\uparrow\right) ,
\end{align}
\begin{align}
\chi^{s}_{\frac{3}{2},\frac{3}{2}}&=\uparrow \uparrow \uparrow , \\
\chi^{s}_{\frac{3}{2},-\frac{3}{2}}&= \downarrow\downarrow\downarrow , \\
\chi^{s}_{\frac{3}{2},\frac{1}{2}}&=\frac{1}{\sqrt3}\left(\uparrow \uparrow \downarrow +\uparrow \downarrow \uparrow +\downarrow \uparrow \uparrow \right) ,\\
\chi^{s}_{\frac{3}{2},-\frac{1}{2}}&=\frac{1}{\sqrt3}\left( \uparrow\downarrow\downarrow+\downarrow\uparrow\downarrow+\downarrow\downarrow\uparrow\right) .
\end{align}
The superscripts, $\mathrm{\sigma}=\rho,\lambda, s , a$, are used to label the symmetry type of the spin wave functions. $s$ stands for the spin wave funcion is completely symmetric and $a$ stands for the wave funcion is completely anty-symmetric. $\rho$ and $\lambda$ are used to label the mixed symmetry. The spin wave functions for the quark-antiquark system are
\begin{align}
\chi^{s}_{1,1}&=\uparrow \uparrow,\quad \chi^{s}_{1,0}=\frac{1}{\sqrt 2}\left(\uparrow \downarrow +\downarrow \uparrow\right),\quad \chi^{s}_{1,-1}=\downarrow \downarrow, \notag \\
\chi^{a}_{0,0}&=\frac{1}{\sqrt 2}\left(\uparrow \downarrow - \downarrow \uparrow\right).
\end{align}

\subsection{The flavor wave functions}
The flavor wave functions for the SU(3) flavor octet baryons, are written as~\cite{LeYaouanc:1988fx}:
\begin{align}
\phi_{\Lambda}^\lambda&=-\frac{1}{2}(sud+usd-sdu-dsu), \\ 
\phi_{\Lambda}^\rho&=\frac{1}{2\sqrt3}(usd+sdu-sud-dsu-2dus+2uds) , 
\end{align}
\begin{align}
\phi_{\Sigma^+}^\lambda &= \frac{1}{\sqrt6}(2uus-suu-usu) ,\\
\phi_{\Sigma^+}^\rho &=\frac{1}{\sqrt2}(suu-usu) ,
\end{align}
\begin{align}
\phi_{\Sigma^0}^\lambda&=\frac{1}{2\sqrt3}(sdu+sud+usd+dsu-2uds-2dus),\\
\phi_{\Sigma^0}^\rho&=\frac{1}{2}(sud+sdu-usd-dsu) ,
\end{align}
\begin{align}
\phi_{\Sigma^-}^\lambda&=\frac{1}{\sqrt6}(sdd+dsd-2dds) ,\\
\phi_{\Sigma^-}^\rho&=\frac{1}{\sqrt 2}(sdd-dsd) ,
\end{align}
\begin{align}
\phi_{\Xi^0}^\lambda&=\frac{1}{\sqrt6}(2ssu-sus-uss) ,\\
\phi_{\Xi^0}^\rho&=\frac{1}{\sqrt2}(sus-uss) ,
\end{align}
\begin{align}
\phi_{\Xi^-}^\lambda&=\frac{1}{\sqrt6}(2ssd-sds-dss) ,\\
\phi_{\Xi^-}^\rho&=\frac{1}{\sqrt2}(sds-dss) .
\end{align}

For the charmed baryons, the flavor wave function are written as ~\cite{Copley:1979wj, Zhong:2007gp, Wang:2017kfr}:
\begin{align}
\phi^{\bar 3}_c&=
\begin{dcases}
\frac{1}{\sqrt2}(ud-du)c, &~ \text{for}~ \Lambda_c^+,  \\  
\frac{1}{\sqrt2}(us-su)c, &~ \text{for}~ \Xi_c^+,\\  
\frac{1}{\sqrt2}(ds-sd)c, &~ \text{for}~ \Xi_c^0,    
\end{dcases}
\\
\phi^{6}_c&=
\begin{dcases}
uuc, &~ \text{for}~ \Sigma_c^{++},  \\  
\frac{1}{\sqrt2}(ud+du)c, &~ \text{for}~ \Sigma_c^{+},\\  
ddc, &~ \text{for}~ \Sigma_c^0, \\
\frac{1}{\sqrt2}(us+su)c, &~ \text{for}~ \Xi_c^{\prime +},\\ 
\frac{1}{\sqrt2}(ds+sd)c, &~ \text{for}~ \Xi_c^{\prime 0},\\
ssc, &~ \text{for}~ \Omega_c^0.     
\end{dcases}
\end{align}

The flavor wave functions of $\pi$ and $K$ are written as\cite{LeYaouanc:1988fx}: 
\begin{align}
\phi_{\pi^+}&=u \bar d, \quad \phi_{\pi^-}=-d\bar u ,\quad \phi_{\pi^0}=\frac{1}{\sqrt2}(u \bar u - d \bar d),\\
\phi_{K^+}&=u\bar s ,\quad \phi_{K^-}=-s \bar u.
\end{align}

\subsection{The spatial wave functions of hadrons}
Ignoring the hyperfine interaction, the Hamiltonian for the three quarks system \cite{Isgur:1978xj, LeYaouanc:1988fx} is expressed as
\begin{align}
H=\sum_{i=1}^3 \frac{\bm p_i^2}{2m_i^2}+\sum_{i\neq j}\frac{1}{2}K|\bm r_i-\bm r_j|^2.
\end{align}
$\bm r_i$ and $\bm p_i$ are the coordinate and momentum for the $i$th quark in the
baryon rest frame.  $K$ is the spring constant that describes the strength of the interaction between constituent quarks. With the Jacobi coordinate the Hamiltonian can be reduced to
\begin{align}
H=\frac{\bm{P}^2}{2 M}+\frac{\bm{p}_\rho^2}{2 m_\rho}+\frac{\bm{p}_\lambda^2}{2 m_\lambda}+\frac{1}{2} m_\rho \omega_\rho^2 \bm{\rho}^2+\frac{1}{2} m_\lambda \omega_\lambda^2 \bm\lambda^2.
\end{align}
where $\bm \rho $ and $\bm \lambda $ are defined as
\begin{align}
\bm \rho&=\frac{1}{\sqrt 2} (\bm r_1-\bm r_2),\\
\bm \lambda&=\sqrt{\frac{2}{3}}\left( \frac{m_1}{m_1+m_2} \bm r_1+\frac{m_2}{m_1+m_2} \bm r_2- \bm r_3 \right) .
\end{align}
$\bm P$ is the total momentum of the baryon. $\bm p_\rho $ and $\bm p_\lambda $ are defined as
\begin{align}
\bm p_\rho    &=m_r \dot{\bm \rho}=\sqrt{2}\left( \frac{m_2}{m_1+m_2}  \bm p_1- \frac{m_1}{m_1+m_2}  \bm p_2\right), \\
\bm p_\lambda &=m_\lambda \dot{\bm \lambda}= \sqrt{\frac{3}{2}}\left( \frac{m_3}{M}  \bm p_1+\frac{m_3}{M}  \bm p_2-\frac{m_1+m_2}{M}  \bm p_3 \right)  .
\end{align}
$M=m_1+m_2+m_3$ is the total mass of the system, $m_\rho$ and $m_\lambda$ are the reduce mass and expressed as
\begin{align}
m_\rho= 2\frac{m_1m_2}{m_1+m_2},~m_\lambda=\frac{3}{2}\frac{m_3(m_1+m_2)}{M} .
\end{align}

In the momentum space, the spatial wave function of the three baryon system can be easily obtained and reads as
\begin{align}
\Psi_{N L L_z}(\bm P,\bm p_\rho,\bm p_\lambda)
&=\delta^3(\bm P-\bm P_c)\sum_m \langle l_\rho,m;l_\lambda,L_z-m|L,L_z \rangle
\notag \\
&\times \psi^{\alpha_\rho}_{n_\rho l_\rho m }(\bm p_\rho)  
\psi^{\alpha_\lambda}_{n_\lambda l_\lambda L_z-m }(\bm p_\lambda),
\end{align}
where
\begin{align}
\psi^\alpha_{nlm}(\bm p)&=(i)^l(-1)^n \left[\frac{2n!}{(n+l+1/2)!} \right]^{\frac12} \frac{1}{\alpha^{l+3/2}}\exp\left({-\frac{\bm p^2}{2\alpha^2}}\right) \notag \\
&\times L_n^{l+1/2}(\bm p^2/\alpha^2)
\mathcal{Y}_{lm}(\bm p) \ .
\end{align}
In the coordinate space, the spatial wave function is expressed as:
\begin{align}
\Psi_{N, L, L_z}\left(\bm{R}_c, \bm{\rho}, \lambda\right)&=\frac{1}{(2 \pi)^{3 / 2}} \exp \left(-i \bm{P} \cdot \bm{R}_c\right)\notag \\
&\times
\sum_m\left\langle l_\rho, m ; l_\lambda, L_z-m \mid L, L_z\right\rangle \psi_{n_\rho l_\rho m}^{\alpha_\rho}(\bm{\rho}) \notag \\
&\times
\psi_{n_\lambda l_\lambda L_z-m}^{\alpha_\lambda}(\bm\lambda),
\end{align}
where 
\begin{align}
\psi_{n l m}^\alpha(\bm{r})&=\left[\frac{2 n!}{(n+l+1 / 2)!}\right]^{1 / 2} \alpha^{l+3 / 2} \exp \left(-\frac{\alpha^2 \bm{r}^2}{2}\right) \notag \\
&\times L_n^{l+1 / 2}\left(\alpha^2 \bm{r}^2\right) \mathcal{Y}_{l m}(\bm{r}) \ .
\end{align}

$\alpha_\rho$ and $\alpha_\lambda$ are the harmonic oscillator strengths. They are expressed as
\begin{align}
\label{eq:arholambda}
\alpha_\rho^2=m_\rho\omega_\rho,~\alpha_\lambda^2=m_\lambda\omega_\lambda ,
\end{align}
with the frequencies of the $\rho$ mode and $\lambda$ mode
\begin{align}
\omega_\rho^2=\frac{3K}{m_\rho},~\omega_\lambda^2=\frac{3K}{m_\lambda}.
\end{align}
With the standard notation, the total principal quantum number $N$ is 
\begin{align} 
N=N_\rho+N_\lambda,
\end{align}
where $N_\rho=(2n_\rho+l_\rho)$ and $N_\lambda=(2n_\lambda+l_\lambda)$ are the principal quantum numbers of the $\rho$ mode and $\lambda$ mode, repectively. 
The total orbital angular-momentum $L$ of a state is obtained by 
\begin{align}
\bm L=\bm l_\rho+\bm l_\lambda.
\end{align}

For the light mesons, only the spatial wave function of the ground state is involved and expressed as:
\begin{equation}
\Psi‘’_{000}(\bm p_1, \bm p_2)= \frac{1}{\pi^{3/4} R^{3/2}}\exp\left[-\frac{(\bm p_1-\bm p_2)^2}{8 R^2}\right],
\end{equation}
where $R$ characterizes the size of mesons.

\subsection{The total wave function of hadrons}
With the above spin, flavor and spatial wave functions, we can construct the total wave functions of the light baryons, which are denoted by $\left|N_6,^{2S+1}N_3,\ N,\ L,\ J \right\rangle$. We also abbreviate $\left|N_6,^{2S+1}N_3,\ N,\ L,\ J \right\rangle$ to $[N_6,^{2S+1}N_3]$ without causing ambiguity. The light baryons wave functions involved in this work read
\begin{widetext}
\begin{align}
|{\bf 56},^28,0,0,\frac{1}{2}\rangle&=\frac{1}{\sqrt2}(\phi_B^\rho \chi^\rho_{S,S_z}+\phi_B^\lambda \chi^\lambda_{S,S_z})\Psi_{0,0,0}  , \\
| {\bf70},^28,1,1,J\rangle&=\sum_{L_z+S_z=J_z}\langle1, L_z;\half, S_z|J J_z\rangle\frac{1}{2}\left[(\phi_B^\rho \chi_{S,S_z}^\lambda+\phi_B^\lambda \chi_{S,S_z}^\rho)\Psi^\rho_{1,1,L_z}+(\phi_B^\rho \chi_{S,S_z}^\rho-\phi_B^\lambda \chi_{S,S_z}^\lambda)\Psi^\lambda_{1,1,L_z} \right] , \\
|{\bf70},^4 8,1,1,J\rangle &= \sum_{L_z+S_z=J_z}\langle1, L_z;\frac{3}{2}, S_z|J J_z\rangle\frac{1}{\sqrt 2}\left[\phi_B^\rho \chi_{S,S_z}^s \Psi^\rho_{1,1,L_z}+ \phi_B^\lambda \chi_{S,S_z}^s\Psi^\lambda_{1,1,L_z} \right] .
\label{eq:lightbaryon}
\end{align}
\end{widetext}
The total wave functions of the ground states and first orbital excitation single chamred baryons are expressed as:
\begin{align}
\Psi_{\bar 3}&=\begin{dcases}
\left |^2 S,\frac{1}{2}^+ \right \rangle 
=\Psi^s_{000}\chi_{S_z}^\rho\phi^{\bar 3}_c \\
\left |^2 P_\lambda,\frac{1}{2}^-\right\rangle =\Psi^\lambda_{11L_z}\chi_{S_z}^\rho\phi^{\bar 3}_c \\
\left |^2 P_\rho,\frac{1}{2}^-\right\rangle =\Psi^\rho_{11L_z}\chi_{S_z}^\lambda\phi^{\bar 3}_c \\
\left |^4 P_\rho,\frac{1}{2}^-\right\rangle 
=\Psi^\rho_{11L_z}\chi_{S_z}^s\phi^{\bar 3}_c
\end{dcases},
\end{align}
\begin{align}
\Psi_{6}&=\begin{dcases}
\left|^2 S,\frac{1}{2}^+\right\rangle
=\Psi^s_{000}\chi_{S_z}^\lambda\phi^{6}_c \\
\left|^2 P_\lambda,\frac{1}{2}^-\right\rangle =\Psi^\lambda_{11L_z}\chi_{S_z}^\lambda\phi^{6}_c \\
\left|^2 P_\rho,\frac{1}{2}^-\right\rangle 
=\Psi^\rho_{11L_z}\chi_{S_z}^\rho\phi^{6}_c \\
\left|^4 P_\lambda,\frac{1}{2}^-\right\rangle 
=\Psi^\lambda_{11L_z}\chi_{S_z}^s\phi^{6}_c
\end{dcases},
\label{eq:sextet}
\end{align}
The Clebsch-Gordan series for the spin and angular-momentum addition 
\begin{align}
\left|{ }^{2 S+1} L_\sigma,\ J^P\right\rangle=
\sum_{L_z+S_z=J_z}\left\langle L L_z, S S_z \mid J J_z\right\rangle \Psi_{N L L_z}^\sigma \chi_{S_z} \phi_c    
\end{align}
 has been omitted for the orbital excitation states and $\sigma=\lambda,\rho,s$ that is also used to label the symmetry type of the wave
function. And the total wave function of pseudoscalar mesons is written as:
\begin{align}
\Phi_{0,0,0} (\bm p_1,\bm p_2)=\delta^3(\bm p_1+\bm p_2-\bm P)\phi_p \chi^a_{0,0}\Psi'_{000}(\bm p_1,\bm p_2).
\end{align}

\end{appendix}


\newpage

\begin{thebibliography}{100}

\bibitem{Wilczek:2004im}
F.~Wilczek,
doi:10.1142/9789812775344{\_}0007
[arXiv:hep-ph/0409168 [hep-ph]].

\bibitem{Shifman:2005wa}
M.~Shifman and A.~Vainshtein,
Phys. Rev. D \textbf{71}, 074010 (2005)
doi:10.1103/PhysRevD.71.074010
[arXiv:hep-ph/0501200 [hep-ph]].

\bibitem{Shifman:2024kfj}
M.~Shifman,
Nucl. Part. Phys. Proc. \textbf{347}, 86-89 (2024)
doi:10.1016/j.nuclphysbps.2024.10.007
[arXiv:2412.05440 [hep-ph]].

\bibitem{Kim:2024tbf}
Y.~Kim, M.~Oka and K.~Suzuki,
Phys. Rev. D \textbf{111}, no.3, 034014 (2025)
doi:10.1103/PhysRevD.111.034014
[arXiv:2411.17803 [hep-ph]].

\bibitem{Ali:2019roi}
A.~Ali, L.~Maiani and A.~D.~Polosa,
Cambridge University Press, 2019,
ISBN 978-1-316-76146-5, 978-1-107-17158-9, 978-1-316-77419-9
doi:10.1017/9781316761465

\bibitem{Chen:2022asf}
H.~X.~Chen, W.~Chen, X.~Liu, Y.~R.~Liu and S.~L.~Zhu,
Rept. Prog. Phys. \textbf{86}, no.2, 026201 (2023)
doi:10.1088/1361-6633/aca3b6
[arXiv:2204.02649 [hep-ph]].

\bibitem{Huang:2023jec}
H.~Huang, C.~Deng, X.~Liu, Y.~Tan and J.~Ping,
Symmetry \textbf{15}, no.7, 1298 (2023)
doi:10.3390/sym15071298

\bibitem{Liu:2019zoy}
Y.~R.~Liu, H.~X.~Chen, W.~Chen, X.~Liu and S.~L.~Zhu,
Prog. Part. Nucl. Phys. \textbf{107}, 237-320 (2019)
doi:10.1016/j.ppnp.2019.04.003
[arXiv:1903.11976 [hep-ph]].

\bibitem{Hosaka:2016pey}
A.~Hosaka, T.~Iijima, K.~Miyabayashi, Y.~Sakai and S.~Yasui,
PTEP \textbf{2016}, no.6, 062C01 (2016)
doi:10.1093/ptep/ptw045
[arXiv:1603.09229 [hep-ph]].

\bibitem{Esposito:2014rxa}
A.~Esposito, A.~L.~Guerrieri, F.~Piccinini, A.~Pilloni and A.~D.~Polosa,
Int. J. Mod. Phys. A \textbf{30}, 1530002 (2015)
doi:10.1142/S0217751X15300021
[arXiv:1411.5997 [hep-ph]].

\bibitem{Klempt:2009pi}
E.~Klempt and J.~M.~Richard,
Rev. Mod. Phys. \textbf{82}, 1095-1153 (2010)
doi:10.1103/RevModPhys.82.1095
[arXiv:0901.2055 [hep-ph]].

\bibitem{Zhu:2004xa}
S.~L.~Zhu,
Int. J. Mod. Phys. A \textbf{19}, 3439-3469 (2004)
doi:10.1142/S0217751X04019676
[arXiv:hep-ph/0406204 [hep-ph]].

\bibitem{Yang:2020atz}
G.~Yang, J.~Ping and J.~Segovia,
Symmetry \textbf{12}, no.11, 1869 (2020)
doi:10.3390/sym12111869
[arXiv:2009.00238 [hep-ph]].

\bibitem{An:2025rjv}
H.~T.~An, S.~Q.~Luo and X.~Liu,
[arXiv:2504.06107 [hep-ph]].

\bibitem{Shi:2021jyr}
P.~P.~Shi, F.~Huang and W.~L.~Wang,
Phys. Rev. D \textbf{103}, no.9, 094038 (2021)
doi:10.1103/PhysRevD.103.094038
[arXiv:2105.02397 [hep-ph]].

\bibitem{Ali:2019clg}
A.~Ali, I.~Ahmed, M.~J.~Aslam, A.~Y.~Parkhomenko and A.~Rehman,
JHEP \textbf{10}, 256 (2019)
doi:10.1007/JHEP10(2019)256
[arXiv:1907.06507 [hep-ph]].

\bibitem{Giannuzzi:2019esi}
F.~Giannuzzi,
Phys. Rev. D \textbf{99}, no.9, 094006 (2019)
doi:10.1103/PhysRevD.99.094006
[arXiv:1903.04430 [hep-ph]].

\bibitem{Maiani:2015vwa}
L.~Maiani, A.~D.~Polosa and V.~Riquer,
Phys. Lett. B \textbf{749}, 289-291 (2015)
doi:10.1016/j.physletb.2015.08.008
[arXiv:1507.04980 [hep-ph]].

\bibitem{Padmanath:2015era}
M.~Padmanath, C.~B.~Lang and S.~Prelovsek,
Phys. Rev. D \textbf{92}, no.3, 034501 (2015)
doi:10.1103/PhysRevD.92.034501
[arXiv:1503.03257 [hep-lat]].

\bibitem{Lee:2009rt}
S.~H.~Lee and S.~Yasui,
Eur. Phys. J. C \textbf{64}, 283-295 (2009)
doi:10.1140/epjc/s10052-009-1140-x
[arXiv:0901.2977 [hep-ph]].

\bibitem{Ebert:2007rn}
D.~Ebert, R.~N.~Faustov, V.~O.~Galkin and W.~Lucha,
Phys. Rev. D \textbf{76}, 114015 (2007)
doi:10.1103/PhysRevD.76.114015
[arXiv:0706.3853 [hep-ph]].

\bibitem{Karliner:2003dt}
M.~Karliner and H.~J.~Lipkin,
Phys. Lett. B \textbf{575}, 249-255 (2003)
doi:10.1016/j.physletb.2003.09.062
[arXiv:hep-ph/0402260 [hep-ph]].

\bibitem{Jaffe:2003sg}
R.~L.~Jaffe and F.~Wilczek,
Phys. Rev. Lett. \textbf{91}, 232003 (2003)
doi:10.1103/PhysRevLett.91.232003
[arXiv:hep-ph/0307341 [hep-ph]].

\bibitem{Selem:2006nd}
A.~Selem and F.~Wilczek,
doi:10.1142/9789812773524{\_}0030
[arXiv:hep-ph/0602128 [hep-ph]].

\bibitem{Garcia-Tecocoatzi:2022zrf}
H.~Garcia-Tecocoatzi, A.~Giachino, J.~Li, A.~Ramirez-Morales and E.~Santopinto,
Phys. Rev. D \textbf{107}, no.3, 034031 (2023)
doi:10.1103/PhysRevD.107.034031
[arXiv:2205.07049 [hep-ph]].

\bibitem{Niu:2020gjw}
P.~Y.~Niu, J.~M.~Richard, Q.~Wang and Q.~Zhao,
Phys. Rev. D \textbf{102}, no.7, 073005 (2020)
doi:10.1103/PhysRevD.102.073005
[arXiv:2003.09323 [hep-ph]].

\bibitem{Ortiz-Pacheco:2024qcf}
E.~Ortiz-Pacheco and R.~Bijker,
[arXiv:2410.09622 [hep-ph]].

\bibitem{Zhou:2023wrf}
Y.~H.~Zhou, W.~J.~Wang, L.~Y.~Xiao and X.~H.~Zhong,
Phys. Rev. D \textbf{108}, no.1, 014019 (2023)
doi:10.1103/PhysRevD.108.014019
[arXiv:2303.13774 [hep-ph]].

\bibitem{Yu:2022ymb}
G.~L.~Yu, Z.~Y.~Li, Z.~G.~Wang, J.~Lu and M.~Yan,
Nucl. Phys. B \textbf{990}, 116183 (2023)
doi:10.1016/j.nuclphysb.2023.116183
[arXiv:2206.08128 [hep-ph]].

\bibitem{Hernandez:2011tx}
E.~Hernandez and J.~Nieves,
Phys. Rev. D \textbf{84}, 057902 (2011)
doi:10.1103/PhysRevD.84.057902
[arXiv:1108.0259 [hep-ph]].

\bibitem{Zhong:2007gp}
X.~H.~Zhong and Q.~Zhao,
Phys. Rev. D \textbf{77}, 074008 (2008)
doi:10.1103/PhysRevD.77.074008
[arXiv:0711.4645 [hep-ph]].

\bibitem{Nagahiro:2016nsx}
H.~Nagahiro, S.~Yasui, A.~Hosaka, M.~Oka and H.~Noumi,
Phys. Rev. D \textbf{95}, no.1, 014023 (2017)
doi:10.1103/PhysRevD.95.014023
[arXiv:1609.01085 [hep-ph]].

\bibitem{Lu:2018utx}
Q.~F.~L{\"u}, L.~Y.~Xiao, Z.~Y.~Wang and X.~H.~Zhong,
Eur. Phys. J. C \textbf{78}, no.7, 599 (2018)
doi:10.1140/epjc/s10052-018-6083-7
[arXiv:1806.01076 [hep-ph]].

\bibitem{Lu:2019rtg}
Q.~F.~L{\"u} and X.~H.~Zhong,
Phys. Rev. D \textbf{101}, no.1, 014017 (2020)
doi:10.1103/PhysRevD.101.014017
[arXiv:1910.06126 [hep-ph]].

\bibitem{Arifi:2021orx}
A.~J.~Arifi, D.~Suenaga and A.~Hosaka,
Phys. Rev. D \textbf{103}, no.9, 094003 (2021)
doi:10.1103/PhysRevD.103.094003
[arXiv:2102.03754 [hep-ph]].

\bibitem{Close:1979bt}
F.~E.~Close,

\bibitem{LeYaouanc:1988fx}
A.~Le Yaouanc, L.~Oliver, O.~Pene and J.~C.~Raynal,

\bibitem{LeYaouanc:1978ef}
A.~Le Yaouanc, O.~Pene, J.~C.~Raynal and L.~Oliver,
Nucl. Phys. B \textbf{149}, 321-342 (1979)
doi:10.1016/0550-3213(79)90244-X

\bibitem{Racah:1942gsc}
G.~Racah,
Phys. Rev. \textbf{62}, 438-462 (1942)
doi:10.1103/PhysRev.62.438

\bibitem{Manohar:1983md}
A.~Manohar and H.~Georgi,
Nucl. Phys. B \textbf{234}, 189-212 (1984)
doi:10.1016/0550-3213(84)90231-1

\bibitem{Zhao:2002id}
Q.~Zhao, J.~S.~Al-Khalili, Z.~P.~Li and R.~L.~Workman,
Phys. Rev. C \textbf{65}, 065204 (2002)
doi:10.1103/PhysRevC.65.065204
[arXiv:nucl-th/0202067 [nucl-th]].

\bibitem{Lee:1957qs}
T.~D.~Lee and C.~N.~Yang,
Phys. Rev. \textbf{108}, 1645-1647 (1957)
doi:10.1103/PhysRev.108.1645

\bibitem{Niu:2025lgt}
P.~Y.~Niu, Q.~Wang and Q.~Zhao,
Phys. Rev. D \textbf{111}, no.9, 093004 (2025)
doi:10.1103/PhysRevD.111.093004
[arXiv:2502.04099 [hep-ph]].

\bibitem{ParticleDataGroup:2024cfk}
S.~Navas \textit{et al.} [Particle Data Group],
Phys. Rev. D \textbf{110}, no.3, 030001 (2024)
doi:10.1103/PhysRevD.110.030001

\bibitem{Zhong:2011ti}
X.~H.~Zhong and Q.~Zhao,
Phys. Rev. C \textbf{84}, 045207 (2011)
doi:10.1103/PhysRevC.84.045207
[arXiv:1106.2892 [nucl-th]].

\bibitem{Niu:2021qcc}
P.~Y.~Niu, Q.~Wang and Q.~Zhao,
Phys. Lett. B \textbf{826}, 136916 (2022)
doi:10.1016/j.physletb.2022.136916
[arXiv:2111.14111 [hep-ph]].

\bibitem{Zhao:2007xb}
Q.~Zhao and F.~E.~Close,
doi:10.1007/978-3-540-85144-8{\_}40
[arXiv:0711.0151 [hep-ph]].

\bibitem{Isgur:1977ef}
N.~Isgur and G.~Karl,
Phys. Lett. B \textbf{72}, 109 (1977)
doi:10.1016/0370-2693(77)90074-0

\bibitem{Isgur:1978xj}
N.~Isgur and G.~Karl,
Phys. Rev. D \textbf{18}, 4187 (1978)
doi:10.1103/PhysRevD.18.4187

\bibitem{Copley:1979wj}
L.~A.~Copley, N.~Isgur and G.~Karl,
Phys. Rev. D \textbf{20}, 768 (1979)
[erratum: Phys. Rev. D \textbf{23}, 817 (1981)]
doi:10.1103/PhysRevD.20.768

\bibitem{Wang:2017kfr}
K.~L.~Wang, Y.~X.~Yao, X.~H.~Zhong and Q.~Zhao,
Phys. Rev. D \textbf{96}, no.11, 116016 (2017)
doi:10.1103/PhysRevD.96.116016
[arXiv:1709.04268 [hep-ph]].

\end{thebibliography}

\end{document}